\documentclass[%
 reprint,
superscriptaddress,
 amsmath,amssymb,
 aps,pre,hyphens,floatfix
]{revtex4-2}

\usepackage[]{amsmath}
\usepackage{mathrsfs}
\usepackage{mathtools} 
\usepackage{enumerate}
\usepackage[usenames,dvipsnames,svgnames,table]{xcolor}
\usepackage{IEEEtrantools}
\usepackage{graphicx}
\usepackage{dcolumn}  
\usepackage{bm}       
\usepackage{array}    
\usepackage{multirow}
\usepackage{soul}
\usepackage[normalem]{ulem}
\usepackage[T1]{fontenc}
\usepackage{subfigure}
\usepackage{diagbox}

\newcommand{\mycomment}[1]{}
\newcommand{\stk}[1]{\ifmmode\text{\sout{\ensuremath{#1}}}\else\sout{#1}\fi}
\begin{document}

\title{Supervised, semi-supervised, and unsupervised learning of
 the Domany-Kinzel model}

\author{Kui Tuo}
\email[]{tuokui@mails.ccnu.edu.cn}
\affiliation{Key Laboratory of Quark and Lepton Physics (MOE) and Institute of Particle Physics,\\ Central China Normal University,
Wuhan 430079, China}
\author{Wei Li}
\email[]{liw@mail.ccnu.edu.cn}
\affiliation{Key Laboratory of Quark and Lepton Physics (MOE) and Institute of Particle Physics,\\ Central China Normal University,
Wuhan 430079, China}

\author{Shengfeng Deng}
\affiliation{School of Physics and Information Technology, Shannxi Normal University, Xi’an 710061, China}

\author{Yueying Zhu}
\affiliation{Research Center of Applied Mathematics and Interdisciplinary Science, Wuhan Textile University, Wuhan 430073, China}

\begin{abstract} 
 The Domany Kinzel (DK) model encompasses several types of non-equilibrium phase transitions, depending on the selected parameters. We apply supervised, semi-supervised, and unsupervised learning methods to studying the phase transitions and critical behaviors of the $(1+1)$-dimensional DK model. The supervised and the semi-supervised learning methods permit the estimations of the critical points, the spatial and temporal correlation exponents, concerning labelled and unlabelled DK configurations, respectively. Furthermore, Principal Component Analysis (PCA) and autoencoder can classify the DK  phases.
 
\end{abstract}

\maketitle

\section{Introduction}

Machine learning (ML) methods have attracted much attention in recent years and have been widely applied to many fields, such as natural language processing \cite{Hinton}, face and image recognition \cite{Parkhi,Krizhevsky}, ecology \cite{Christin}, economics and finance \cite{Gogas}, data mining and analysis \cite{Ge}, and electronic games \cite{Justesen}, etc. Recently, reinforced learning methods even have solved the previously unfathomable go games (AlphaGo) \cite{Silver}. Machine learning methods also show their great advances in boosting multidisciplinary problem solving, especially when the related tasks are data-driven optimization problems. 

In the physics realm, some major progresses have also been made with machine learning methods. For example, quantum computing has been combined with machine learning to develop the field of quantum machine learning: In Ref.~\cite{Benedetti} the quantum annealer has been used to sample the Boltzmann distribution, and Ref.~\cite{Amin} and Ref.~\cite{Farhi} have studied the quantum Boltzmann machine and constructed quantum neural network, respectively. In high energy physics, a machine learning classifier has been constructed to search for new particles of unknown masses~\cite{Baldi}, using parameterized networks to simplify the training process and enhance the learning performance. The quantum chromodynamics (QCD) phase transition has also been studied by using deep convolutional neural networks~\cite{Pang}. More recently, graph neural networks (GNNs) combined with a HaarPooling operation have been applied to extracting the features of quark-gluon tagging\cite{ma2022jet}, which can enhance the accuracy of quark-gluon tagging, as compared to the weakly supervised learning method proposed earlier \cite{Dery}. In astrophysics, the machine learning package astroML has been developed ~\cite{VanderPlas}, and machine learning methods have also been utilized to boost cosmological and astrophysics process simulations \cite{Villaescusa}.  

Important breakthroughs have also been made in learning different phases of matters. The seminal work by Carrasquilla and Melko in 2016 have demonstrated that the ferromagnetic and paramagnetic phases of the classical Ising model can be classified based on supervised machine learning methods  \cite{Carrasquilla}, permitting the identification of the critical points and the spatial correlation exponents. This work has since triggered great interest in the application of machine learning methods to the studies of various types of phase transitions. Regardless of the complexity of the target problem, the versatility of machine learning methods allows for learning more complex phases of 3-dimensional Ising model \cite{Zhang}, or phases with non-local and topological (Kosterlitz-thouless) properties in percolation, XY and generalized XY models \cite{ZhangW,Beach,rodriguez2019identifying}, or even phases of non-equilibrium matters \cite{Venderley}, such as many-body localized and topological phases, and the non-equilibrium phase transitions in the directed percolation (DP) \cite{Li}. As in Ref.~\cite{Carrasquilla}, it has been repeatedly demonstrated that one can estimate the critical points and the spatial correlation exponents, which further enhances the possibility for obtaining the entire phase diagram that is consistent with theory \cite{Venderley}. 

In this work, we study the phase transitions of the (1+1)-dimensional Domany-Kinzel (DK) model by machine learning techniques. As will be shown in the next section, the DK model is controlled by two parameters. Along the transition line, the model characterizes several types of phase transitions. Hence the DK model provides an excellent test bed for comparing the capabilities of different learning methods. To that end,  supervised, semi-supervised, and unsupervised learning methods will be applied to each type of phase transition. For supervised learning, the respective critical points and critical exponents are estimated.  We also propose a new semi-supervised learning method, in which only half probability data of training set with respect to test set are labelled. The trained neural network can then predict the order parameter of the unlabelled DK model and the corresponding critical points. In addition, two unsupervised learning methods, Principal Component Analysis (PCA) and autoencoder can classify the DK phases.

The remainder of this paper is organized as follows: In Sec.~\ref{sec:DK}, we briefly introduce the DK model. Sec.~\ref{sec:DK Supervised} presents the supervised learning of (1+1)-dimensional DK model, in which the critical points and the correlation length and correlation time exponents are estimated. Sec.~\ref{sec:DK semi-supervised} gives the semi-supervised learning results of (1+1)-dimensional DK model. Sec.~\ref{sec:DK Unsupervised} is about the unsupervised learning results of (1+1)-dimensional DK model, via autoencoder and PCA. Sec.~\ref{sec:Summary} summarizes the main findings of this work. 

\section{The Domany-Kinzel model \label{sec:DK}}

The Domany-Kinzel (DK) \cite{domany1984equivalence,Hinrichsen} model is a stochastic cellular automaton that exhibits non-equilibrium active-to-absorbing type phase transition, controlled by two parameters. In (1+1) dimensions, the model is defined on a one-dimensional array, on which site $s_i$ can be either occupied ($s_i=1$) or empty ($s_i=0$). As illustrated in Fig.~\ref{fig:DKillu}, the state of each site is then updated with time with respect to its nearest neighbors according to the following rule:

\begin{figure}[t]
  \centering
    \includegraphics[width=1\columnwidth]{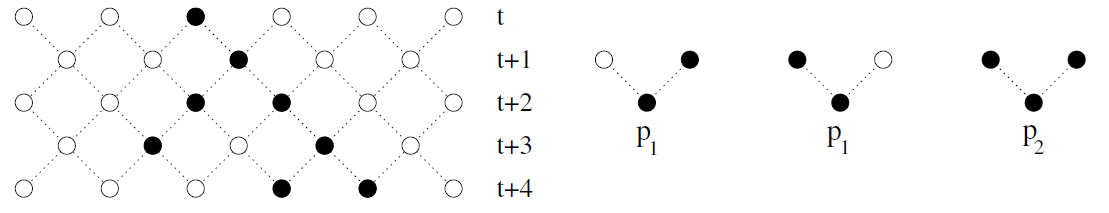}
    \caption{The (1+1)-dimensional Domany-Kinzel model \cite{Hinrichsen}. Occupied sites are marked by black circles. The state $s_{i,t+1}$ of a given site $i$ at time $t+1$ depends on the states of its left and right neighbours ($s_{i-1,t}$, $s_{i+1,t}$) at time step $t$.}
   \label{fig:DKillu}
\end{figure}

\begin{equation}
 s_i(t+1) = \begin{cases}
 1&\text{if }  s_{i-1}(t)\not=s_{i+1}(t)\qquad\,\text{and} \  r_i(t)<p_1\\ 
 1&\text{if }  s_{i-1}(t)=s_{i+1}(t)=1\ \text{and} \ r_i(t)<p_2\\ 
 0&\text{otherwise}\,,
\end{cases}
\label{eqs:DKrule}
 \end{equation}
where $0\le r_i(t) \le 1$ is a random number generated from a uniform distribution, and $0\le p_1 \le 1$ and $0\le p_2 \le 1$ are two probabilities used to control the phases of the model.

From the above rule, one can easily imagine that unless all the sites are initially occupied, given the probability $p_2$, if $p_1$ is too small, the proportion of the occupied sites will decrease until only empty sites remain, whereas, for large enough $p_1$ value, the array will become increasingly occupied until a saturated density is reached. Once the system evolves into a fully empty state, there is no way for it to get out of that state. Hence the model displays an active-to-absorbing phase transition.

\begin{figure}[ht]
  \centering
    \includegraphics[width=0.8\columnwidth]{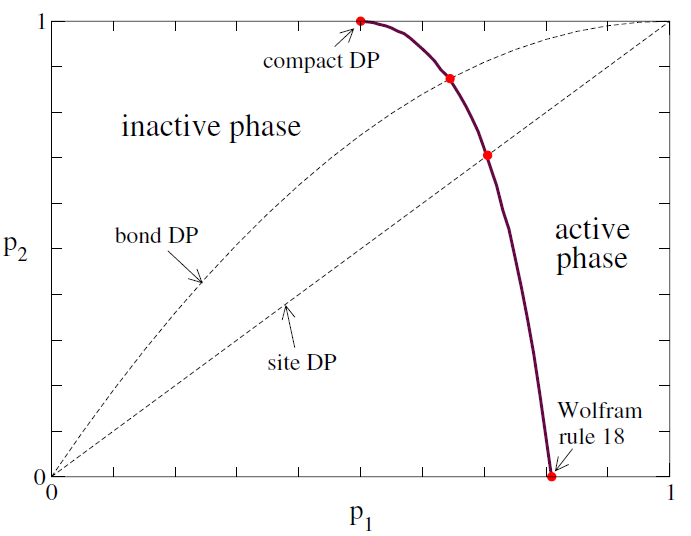}
    \caption{Phase diagram of the (1+1)-dimensional Domany-Kinzel model \cite{Hinrichsen}. Bond directed percolation corresponds to the line $p_2$=$p_1$($2-p_1$). Site directed percolation is obtained for $p_1$ = $p_2$. For $p_2$=0, it is equivalent to Wolfram rule 18. For $p_2$=1, it is a different universal scaling behaviour called compact directed percolation.}
   \label{fig:DKphase}
\end{figure}

\begin{figure}[!htbp]
\setlength{\tabcolsep}{0pt}
\centering
\begin{tabular}{cccc}
\includegraphics[width=0.245\columnwidth]{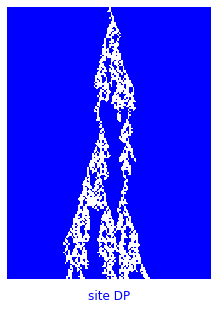} &
\includegraphics[width=0.245\columnwidth]{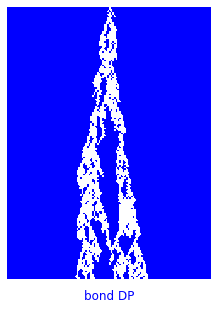} &
\includegraphics[width=0.245\columnwidth]{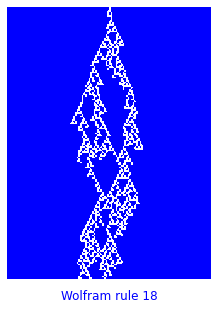} &
\includegraphics[width=0.245\columnwidth]{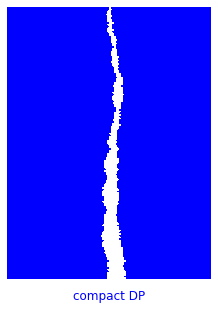}
\end{tabular}
\caption{Critical clusters of DK model generated from a single active seed.}
\label{fig:DKseed}
\end{figure}

As depicted in Fig.~\ref{fig:DKphase} for the phase diagram, there is a transition line $(p_{1c}, p_{2c})$ separating the active phase and the absorbing one. Depending on the location of the parameters, the DK model includes bond and site DP as special cases. Bond DP corresponds to the line $p_2$= $p_1$($2-p_1$), while site DP is obtained for $p_1$ = $p_2$. For $p_2$=0, it is equivalent to Wolfram rule 18, and for $p_2$=1, since an empty site is \textit{guaranteed} to be filled as long as both its neighbors are occupied, the generated clusters become compact, giving rise to a different universal scaling behavior called the compact directed percolation (CDP). The CDP is different from the bond DP, the site DP, and the Wolfram rule 18, as the latter ones all belong to the DP universality \cite{janssen1981nonequilibrium,grassberger1982phase}. This is exemplified by the critical clusters of the DK model generated from a single active seed shown in Fig.~\ref{fig:DKseed}, in which the compact pattern of the CDP cluster is quite distinctive from those of the rest ones. 

The order parameter of the DK is defined as the density of active sites
\begin{equation}
 \rho (t)=\big\langle \frac{1}{N}\sum_{i} s_i(t)\big\rangle\,.
\label{eqs:orderp}
 \end{equation}
We first consider the case of an infinite system. In the active phase, $\rho (t)$ decays initially and eventually saturates at some stationary value $\rho_{stat}$ that varies according to a power law characterized by the exponent $\beta$ near the critical point:
\begin{equation}
\rho_{stat} \sim (p-p_c)^{\beta}\,.
 \end{equation}

\section{Supervised learning of the Domany-Kinzel model \label{sec:DK Supervised}}

The inputs for the learning machines are just raw configurations generated from Monte Carlo(MC) simulations of the (1+1)-dimensional DK model. According to the phase diagram depicted in Fig.~\ref{fig:DKphase}, each type of phase transition of the DK model can be controlled by varying the probability $p_1$. We henceforth denote it as $p$ for simplicity. The generated configurations are split into the training set and the test one. In the training set, each configuration is labelled according to the probability $p$ that generated that configuration. The labelling is a prerequisite for the supervised learning.
If the probability $p$ of a configuration $\mathbf{x}_i$ is smaller than the critical probability $p_c$, it is in the absorbing phase and labelled as $\mathbf{y}_i=(0, 1)$; otherwise, if $p > p_c$, it is in the active phase and labelled as $\mathbf{y}_i=(1, 0)$.

\begin{figure}[t]
  \centering
    \includegraphics[width=1\columnwidth]{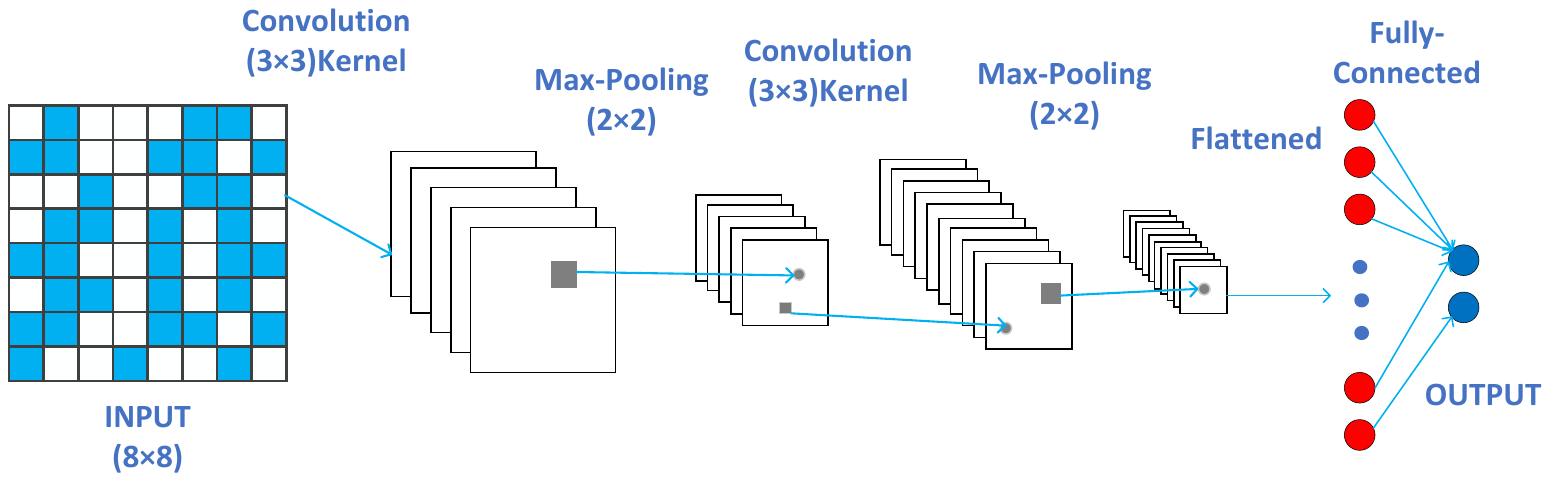}
    \caption{Schematic structure of CNN.}
   \label{fig:CNN}
\end{figure}

For supervised learning, we apply the convolutional neural network (CNN) as illustrated in Fig.~\ref{fig:CNN}, on which, sigmoid activation function is used for the convolution and the pooling layers, and softmax activation function is taken in the output layer, producing a binary classification output. For a certain test configuration $\mathbf{x}_i(p)$ fed into the neural network, one output layer signifies the probability that the configuration belongs to the active phase $P_1(\mathbf{x}_i(p))$, and the other output layer signifies the probability that the configuration belongs to the absorbing phase $P_0(\mathbf{x}_i(p))$.

\begin{figure}[!htbp]
\setlength{\tabcolsep}{0pt}
\centering
\begin{tabular}{cccc}
\includegraphics[width=0.245\columnwidth]{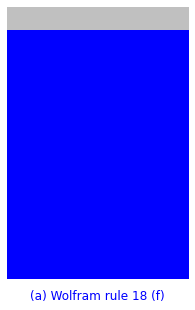} &
\includegraphics[width=0.245\columnwidth]{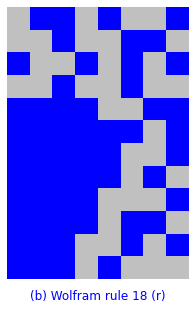} &
\includegraphics[width=0.245\columnwidth]{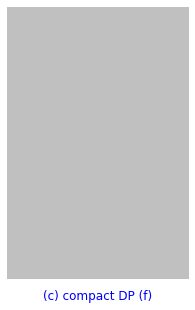} &
\includegraphics[width=0.245\columnwidth]{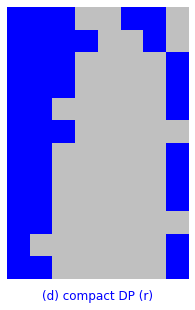}
\end{tabular}
\caption{(a) Cluster of Wolfram rule 18 generated from fully occupied active seeds. (b) Critical cluster of Wolfram rule 18 generated from randomly occupied active seeds. (c) Cluster of CDP generated from fully occupied active seeds. (d) Critical cluster of CDP generated from randomly occupied active seeds. Gray colour represents occupied sites, and blue ones represent empty sites. }
\label{fig:Wolfram_CDP}
\end{figure}

\begin{figure}[!htbp]
\setlength{\tabcolsep}{0pt}
    \centering
    \begin{tabular}{cc}
    \includegraphics[width=0.49\columnwidth]{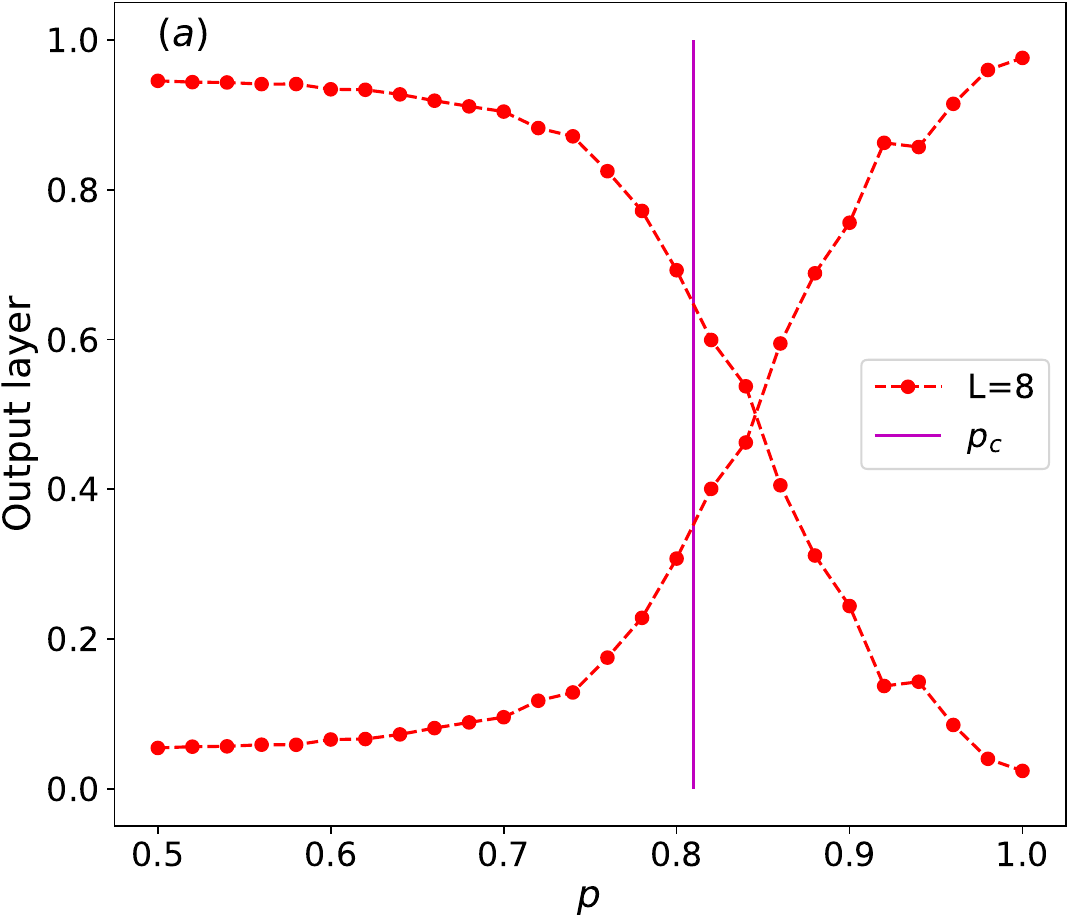} &
    \includegraphics[width=0.49\columnwidth]{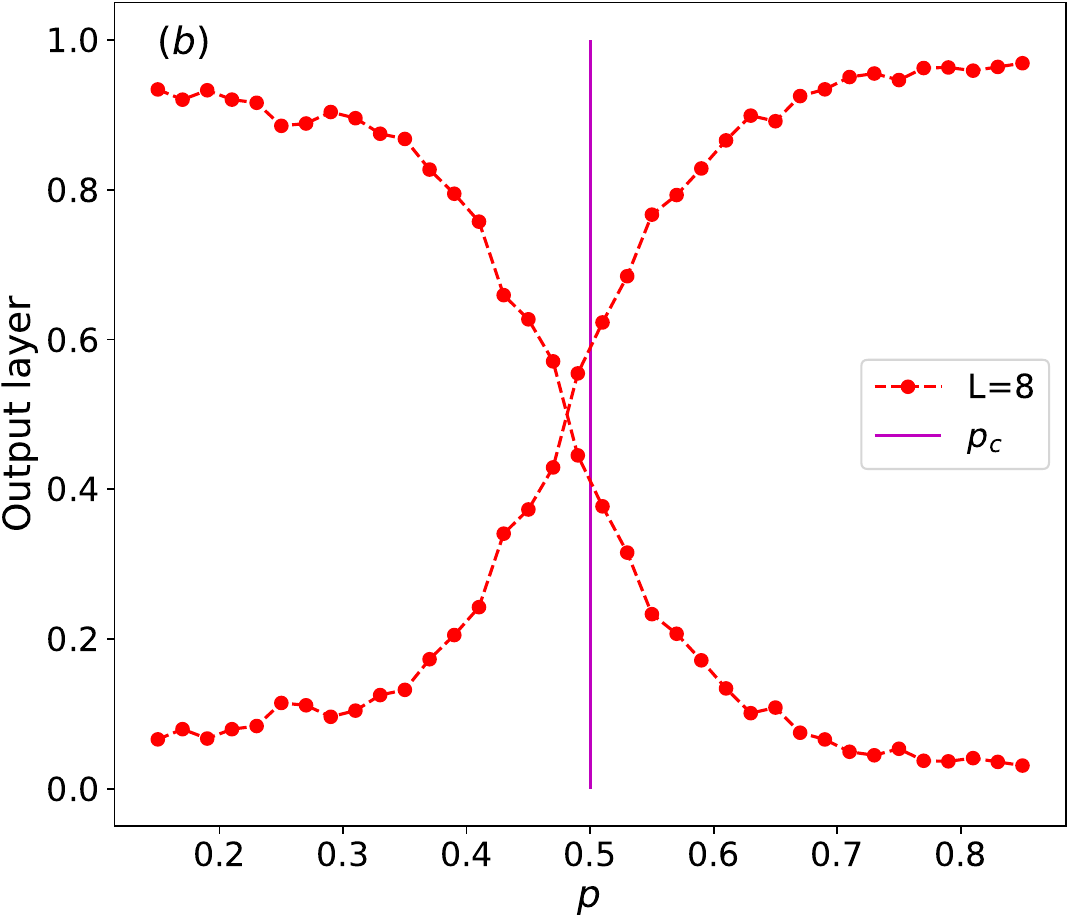}\\
    \includegraphics[width=0.49\columnwidth]{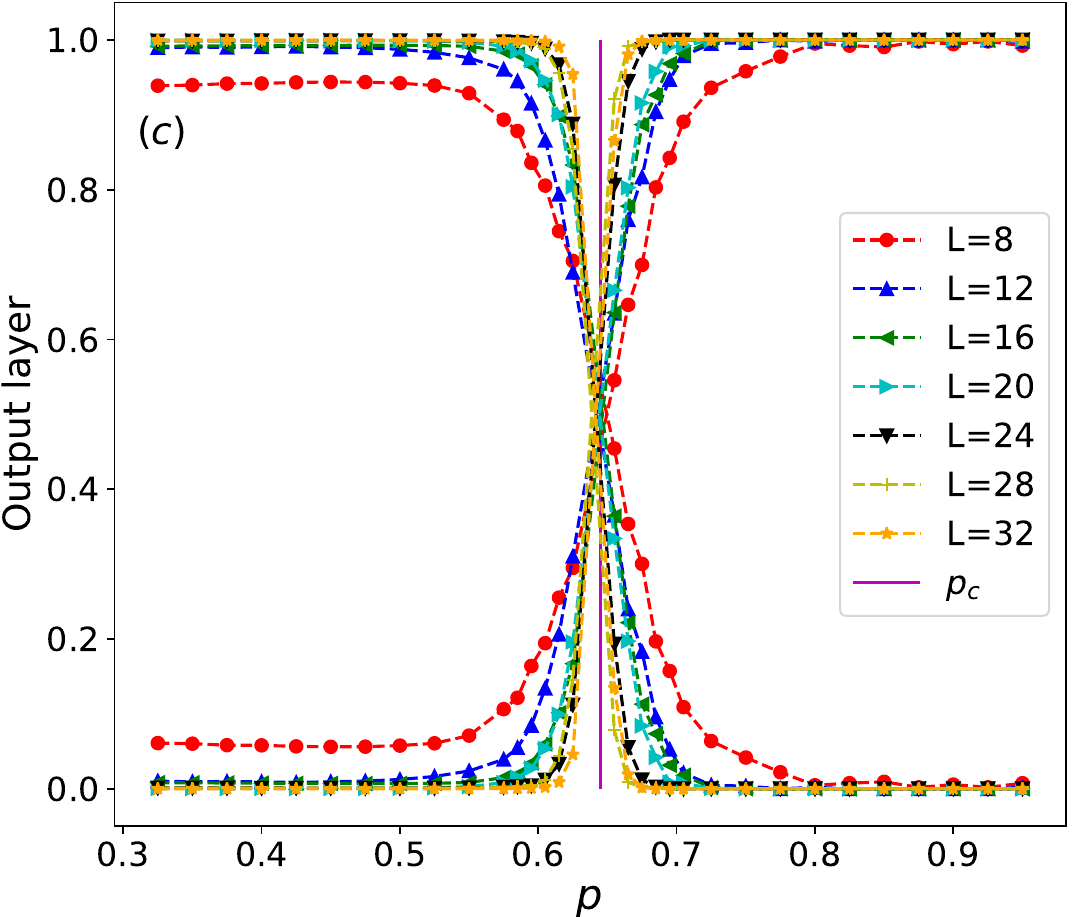} &
    \includegraphics[width=0.49\columnwidth]{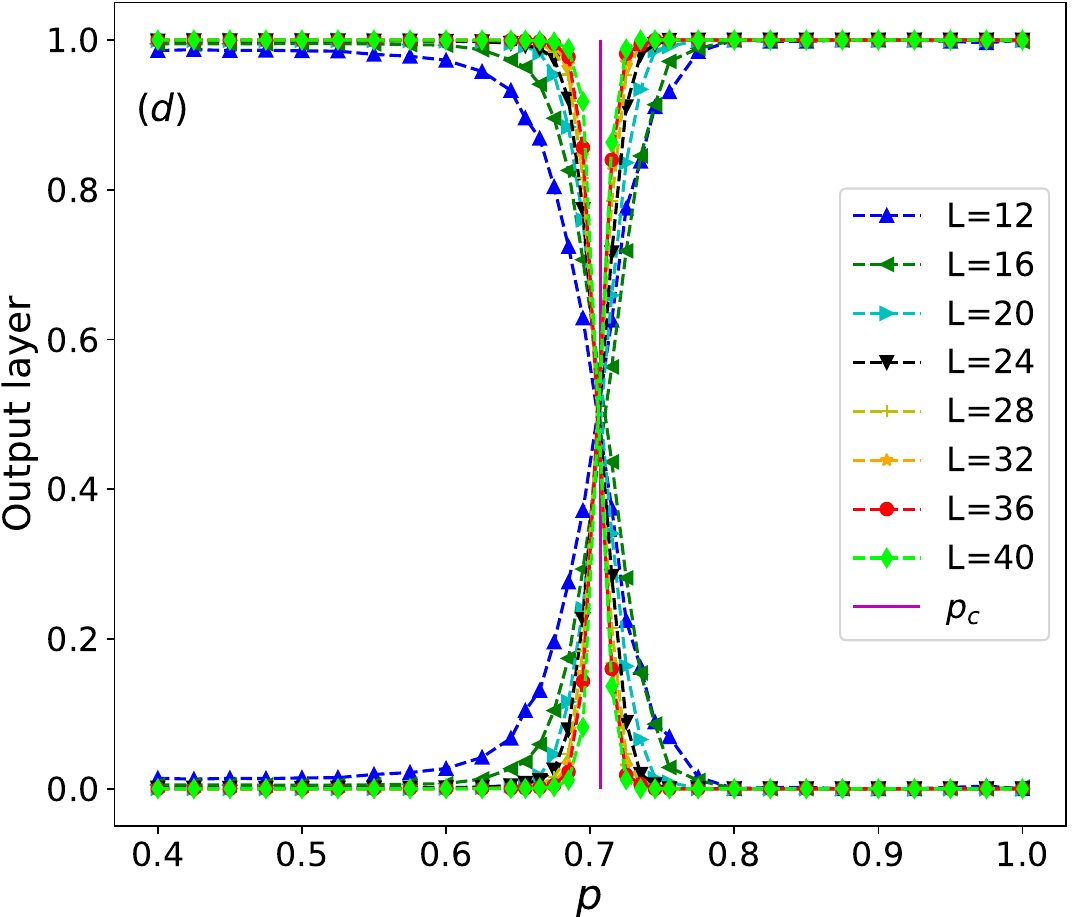}\\
    \includegraphics[width=0.49\columnwidth]{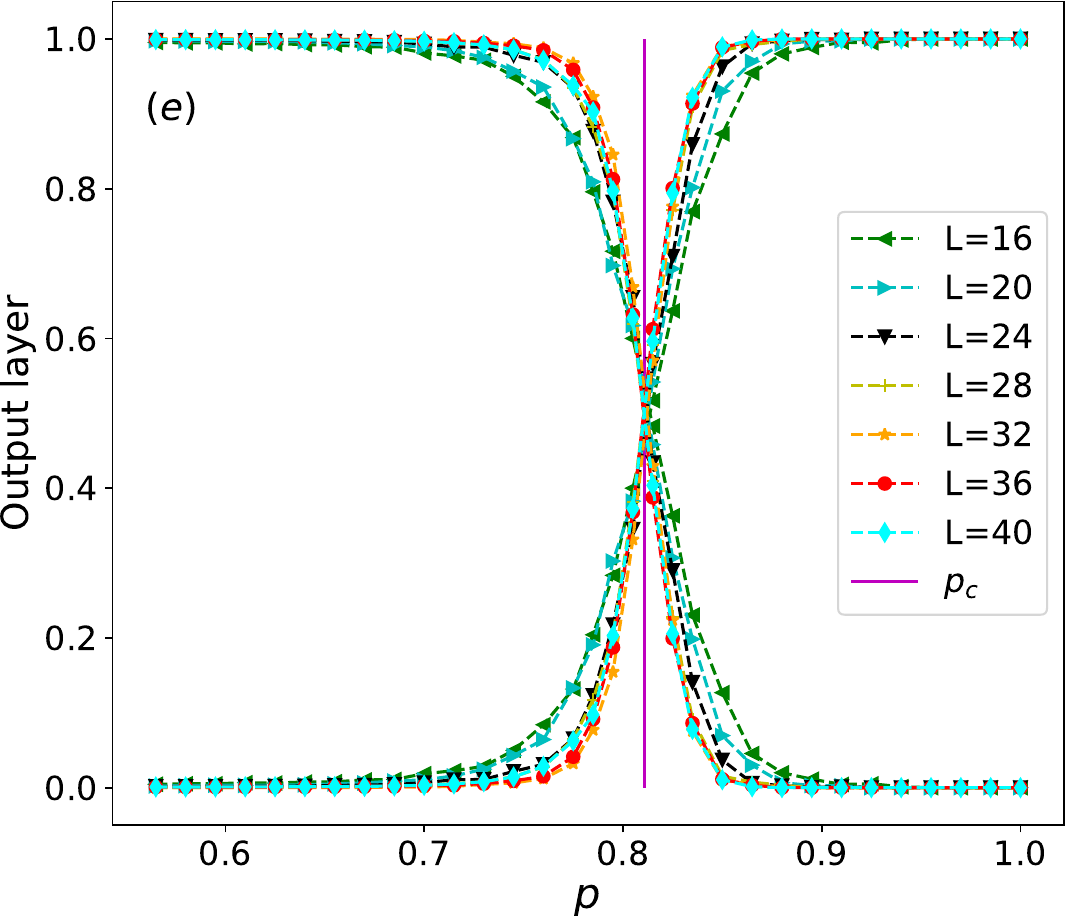} &
    \includegraphics[width=0.49\columnwidth]{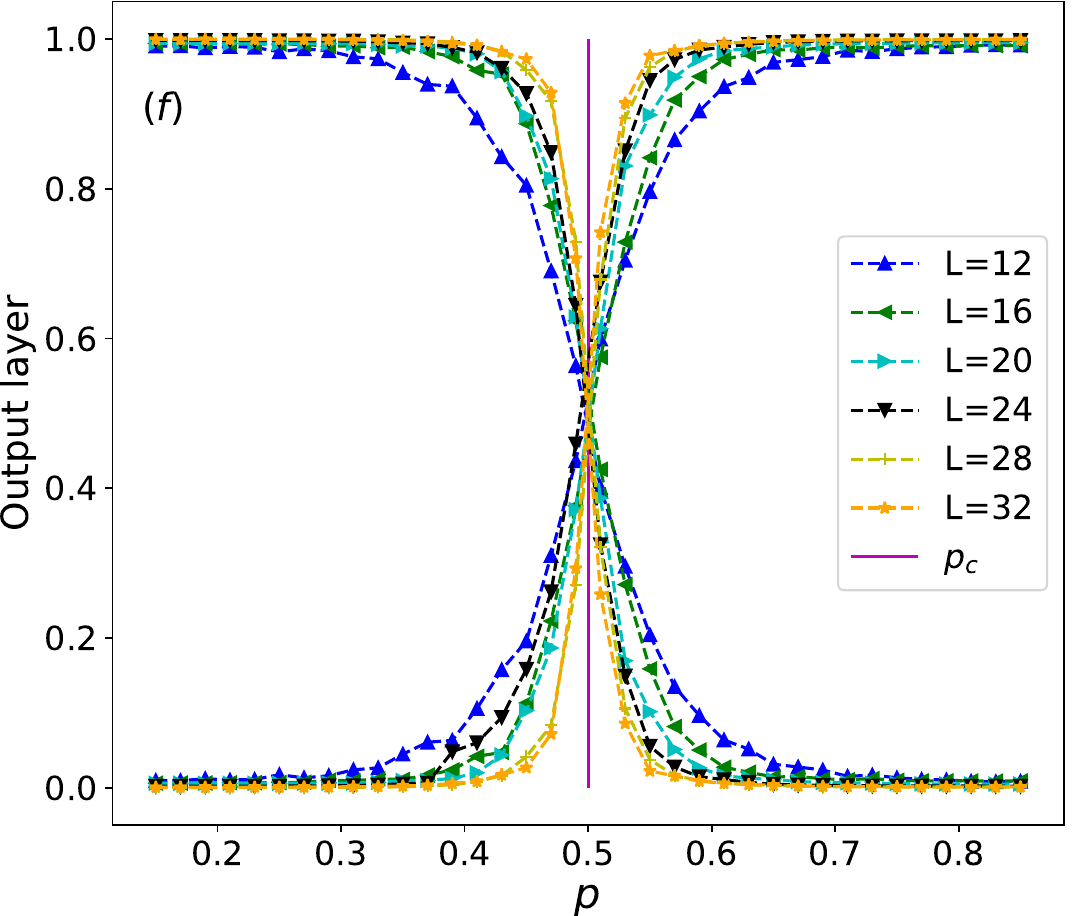}
    \end{tabular}
    \caption{Averaged learning results of CNN output layers for (a) the Wolfram rule 18 ($L=12$, $t=60$); (b) the CDP($L=8$, $t=64$); (c) the bond DP ($L=8$, 12, 16, 20, 24, 28, 32; t=50, 60, 80, 115, 152, 194, 240); (d) the site DP ($L=12$, 16, 20, 24, 28, 32, 36, 40; $t=60$, 80, 115, 152, 194, 240, 288, 340);(e) the Wolfram rule 18 ($L=16$, 20, 24, 28, 32, 36, 40; $t=80$, 115, 152, 194, 240, 288, 340); and (f) the CDP ($L=12$, 16, 20, 24, 28, 32; $t=145$, 256, 400, 576, 784, 1024).} 
    \label{fig:supervised} 
\end{figure}

Since learning machines try to learn features of configuration images of relatively small system sizes and of different phases, it is customary to make use of configurations obtained from fully occupied or randomly occupied (e.g.~with a probability of 0.5) initial states instead of starting from a single active seed as shown in Fig.~\ref{fig:DKseed}, as the latter leaves a large proportion of the sites empty at initial stages.  Here, randomly occupied initial states are more preferable because for the Wolfram rule 18 ($p_2=0$) and the CDP ($p_1=1$), fully occupied initial states only result in trivial absorbing and active states, as illustrated in Fig.~\ref{fig:Wolfram_CDP} (a) and (c), respectively, regardless of how the other parameter ($p_1$ for the Wolfram rule 18 and $p_2$ for the CDP) is chosen. 

Simulation times $t$ are typically selected with respect to the characteristic time $t_f$. On a finite lattice of non-equilibrium phase transition, there is always a non-vanishing probability of reaching the absorbing configuration, finite-size effects set in after a typical time $t_f$ that grows with the system size as $t_f \sim L^z$. For the DP universality class, $z=1.58$ \cite{jensen1996low}, while for the CDP, $z=2$ \cite{domany1984equivalence}.

The configuration images are of $L \times (t+1)$ dimension. From initial states with randomly occupied sites, for each probability $p$, 1700 labelled configurations are generated for the training set and another 500 configurations for the test set. The CNN output layers are eventually averaged over the test set, giving $P_{0|1}(p)=\frac{1}{500}\sum_{i=1}^{500} P_{0|1}(\mathbf{x}_i(p))$.

To examine the proper system size, we started with $L=12$ and $L=8$ for the Wolfram rule 18 and the CDP, which give rise to the relatively low accuracy values of $89.17\% $ and $89.8\% $, respectively (c.f.~Fig 6(a) and (b)). Therefore, larger sizes will be taken in what follows. Fig.~\ref{fig:supervised}(c) and (d) show the averaged results for the output layers from the above-trained CNN  with respect to bond and site DP, respectively. The critical points can be estimated from the crossing point of the two output layers, which is typically around $P_0(p_c)=P_1(p_c)\approx 0.5$. With $L=32$, we hence estimate $p_c=0.643\pm 0.02$ for the bond DP. With $L=40$, we estimate $p_c=0.699\pm 0.02$ for the site DP. These estimations yield accuracy of $99.54\% $ for the bond DP critical point, and accuracy of $99.37\%$ for the site DP; see the accuracy data in Table~\ref{table:CNN accuracy_L}. Therefore, even though the employed system sizes are relatively small, supervised learning by CNN still allows us to classify the two phases and estimate the associated critical points quite well. 

For Fig.~\ref{fig:supervised}(e) and (f), with $L=40$, we estimate $p_c=0.796\pm 0.02$ for the Wolfram rule 18. With $L=32$, we estimate $p_c=0.498\pm 0.02$ for the CDP. These estimations yield accuracy of $97.05\% $ for the Wolfram rule 18 critical point, and accuracy of $98.28\%$ for the CDP.

\begin{table*}[htbp]
\centering
\caption{The accuracy values of the trained CNNs with respect to different system sizes $L$ for (a) the bond DP, (b) the site DP, (c) the Wolfram rule 18, and (d) the CDP.}
\begin{tabular}{cccccccccc}
\hline\hline
size $L$ & 8 & 12 & 16 & 20 &24 & 28 & 32 & 36 & 40 \\
\hline
bond DP& $92.97\%$ & $96.05\%$ & $ 97.37\% $ & $ 97.71\% $ & $ 99.11\% $ & $ 99.37\% $ & $99.54\%$ & $-$ & $-$  \\
\hline
site DP & $ 90.64\%$ & $ 95.02\%$ & $ 96.07\%$ & $ 97.46\%$ & $ 98.12\%$ & $ 98.83\%$ & $ 99.07\%$ & $ 99.10\%$ & $ 99.37\%$\\
\hline
Wolfram rule 18 & $ 86.72\%$ & $ 89.17\%$ & $ 94.06\%$ & $ 95.31\%$ & $ 96.22\%$ & $ 96.2\%$ & $ 96.78\%$ & $ 96.91\%$ & $ 97.05\%$\\
\hline
CDP& $ 89.80\%$ & $ 94.41\%$ & $ 95.90\%$ & $ 96.70\%$ & $ 97.56\%$ & $ 98.03\%$ & $ 98.39\%$ & $-$ & $-$\\
\hline\hline
\end{tabular}
\label{table:CNN accuracy_L}
\end{table*}

\begin{figure}[t]
  \centering
    \includegraphics[width=0.8\columnwidth]{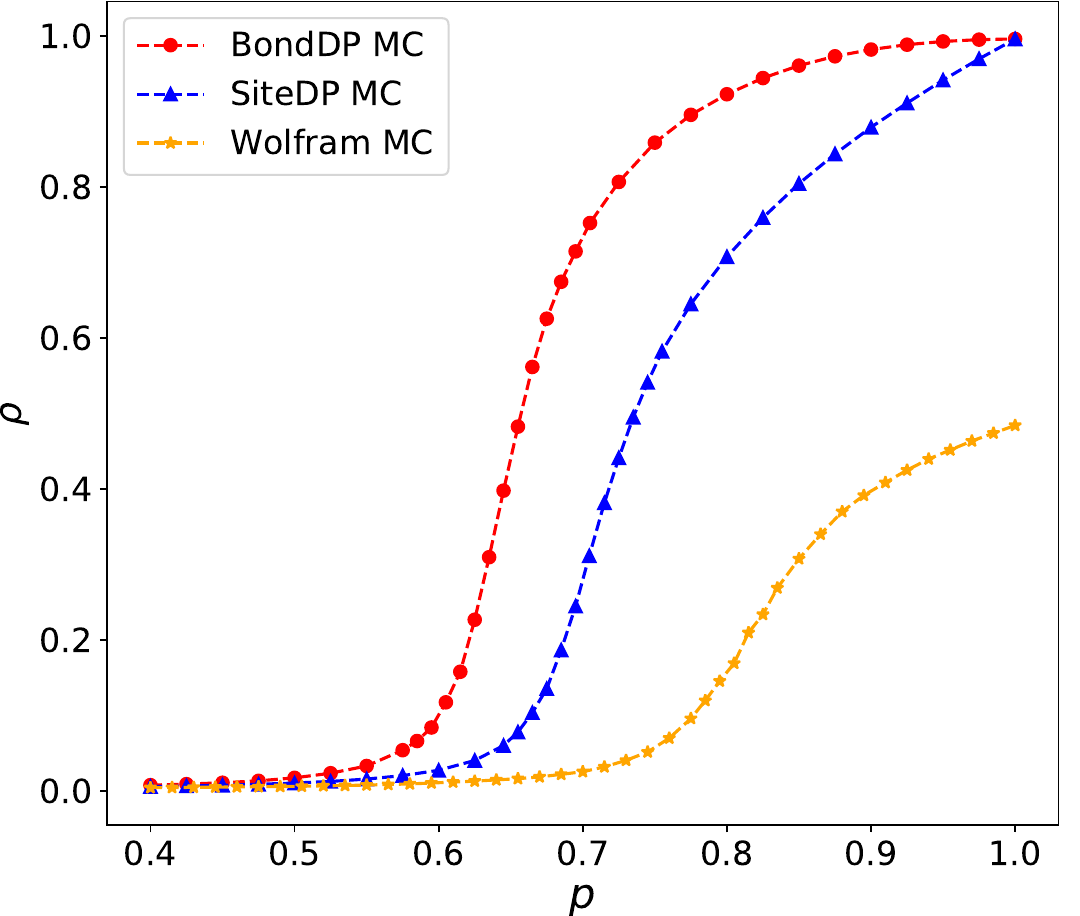}
    \caption{With $L=32$, $t=240$, comparing the particle number density of the bond DP, site DP and  Wolfram rule 18.}
   \label{fig:BondDP_site_Wolfram_MC}
\end{figure}

As one can notice from Table.~\ref{table:CNN accuracy_L}, the accuracy values for the bond DP, the site DP, and the Wolfram rule 18 display a decreasing trend amongst them for each studied system size. As shown in Fig.~\ref{fig:BondDP_site_Wolfram_MC}, the particle number densities of the three models at the same $p$ are actually different and display the same trend. Since these three variants all belong to the DP universality class, here we observe that the non-universal “lacunarity” property of clusters, associated with the particle number density, still affects the learning accuracy, although one may just want to probe the same universality properties. 

The features of non-equilibrium phase transitions such as absorbing phase transitions are encoded in the correlations within the spatial configurations and their dynamical evolution. Approaching the critical point, the spatial correlation length $\xi _\bot$ and the temporal correlation length $\xi _\Vert$ diverge as
\begin{equation}
\xi_\bot \sim \vert p-p_c\vert^{-\nu_\bot} \quad \text{and} \quad \xi_\Vert \sim \vert p-p_c \vert^{-\nu_\Vert},
\end{equation}
where $\nu _\bot$ and $\nu _\Vert$ are spatial and temporal correlation exponents, respectively, and $\xi _\Vert \sim  \xi _\bot^z$, with $z=\nu_\Vert/ \nu_\bot$ being the dynamical exponent.
For finite systems simulated within finite times, by noting that $\xi_\bot \sim \vert p-p_c\vert^{-\nu_\bot} \sim L$, $\xi_\Vert \sim \vert p-p_c \vert^{-\nu_\Vert} \sim t$, one sees that $x=(p-p_c)L^{1/\nu_\bot}$ and $y=(p-p_c)t^{1/{\nu_{\Vert}}}$ are dimensionless quantities, so the functions $\hat{P}_{0|1}(x)$ and $\hat{P}_{0|1}(y)$ are scaling functions of $x$ and $y$, respectively. Hence, it is possible to estimate $\nu_\bot$ and $\nu_\Vert$ by performing data collapse techniques.

\begin{figure}[!htbp]
\setlength{\tabcolsep}{0pt}
    \centering
    \begin{tabular}{cc}
    \includegraphics[width=0.49\columnwidth]{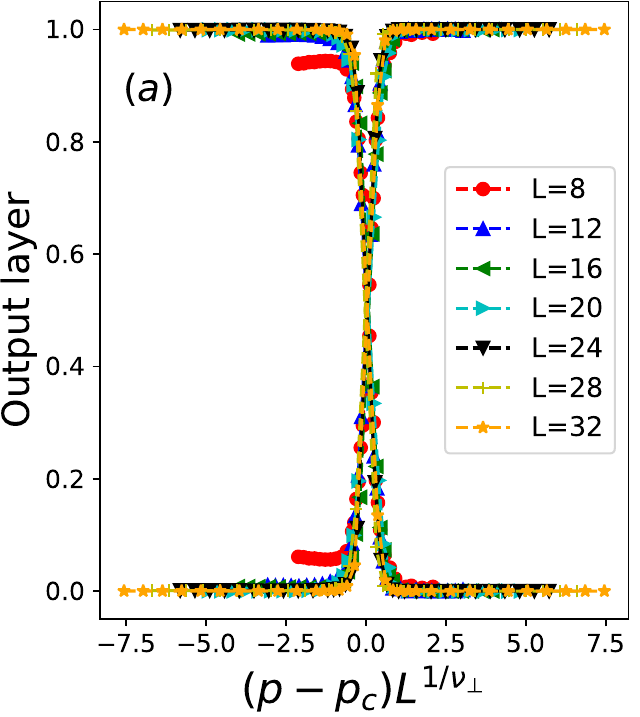} &
    \includegraphics[width=0.49\columnwidth]{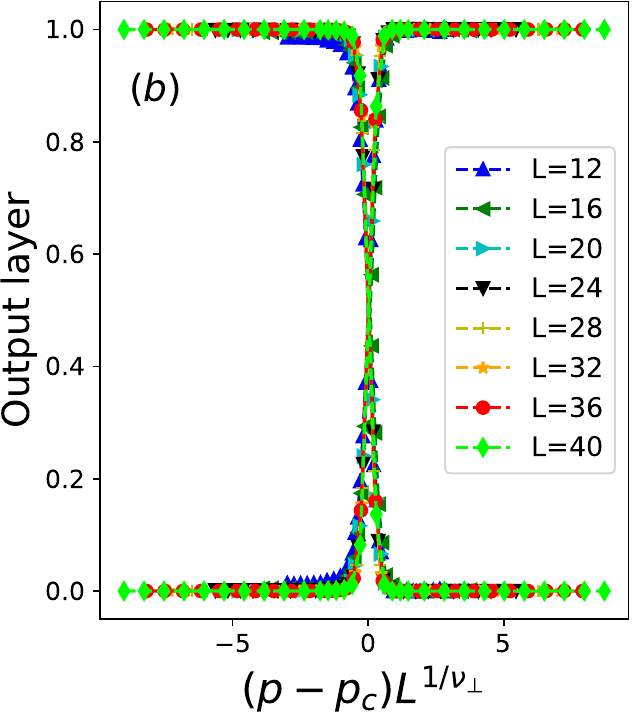} \\
    \includegraphics[width=0.49\columnwidth]{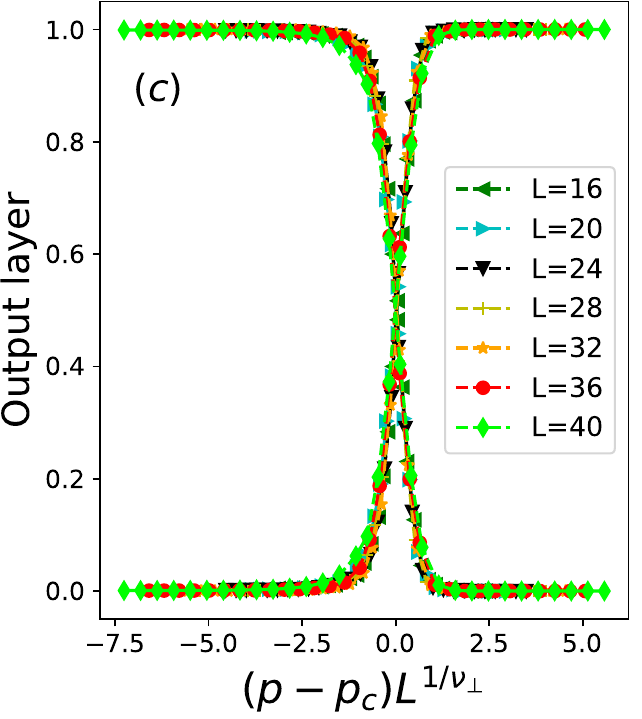} &
    \includegraphics[width=0.49\columnwidth]{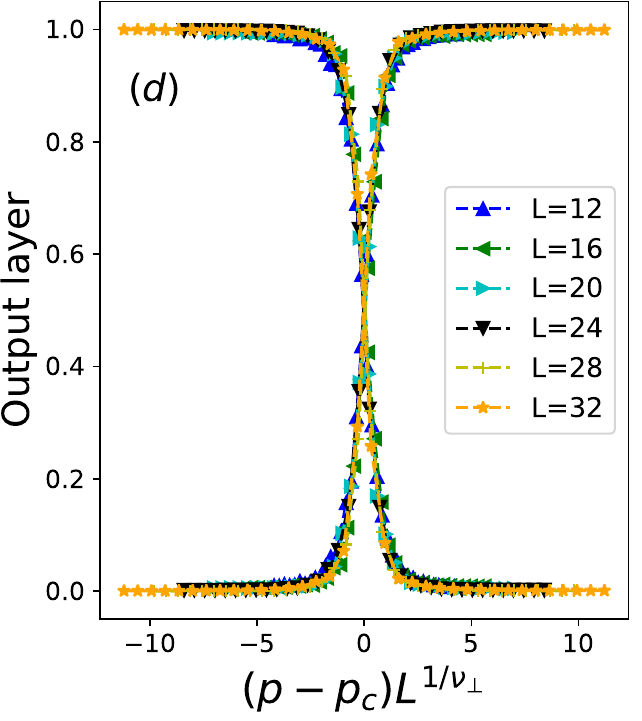}
    \end{tabular}
    \caption{CNN outputs results as a function of $(p-p_c)L^{1/{\nu_{\bot}}}$ for (a) the bond DP ($L=8$, 12, 16, 20, 24, 28, 32; $t=50$, 60, 80, 115, 152, 194, 240), (b) the site DP ($L=12$, 16, 20, 24, 28, 32, 36, 40; $t=60$, 80, 115, 152, 194, 240, 288,340), (c) the Wolfram rule 18 ($L=16$, 20, 24, 28, 32, 36, 40; $t=80$, 115, 152, 194, 240, 288, 340), (d) the CDP ($L=12$, 16, 20, 24, 28, 32; $t=145$, 256, 400, 576, 784, 1024).}
    \label{fig:collapse1}
\end{figure}

\begin{figure}[!htbp]
\setlength{\tabcolsep}{0pt}
    \centering
    \begin{tabular}{cc}
    \includegraphics[width=0.49\columnwidth]{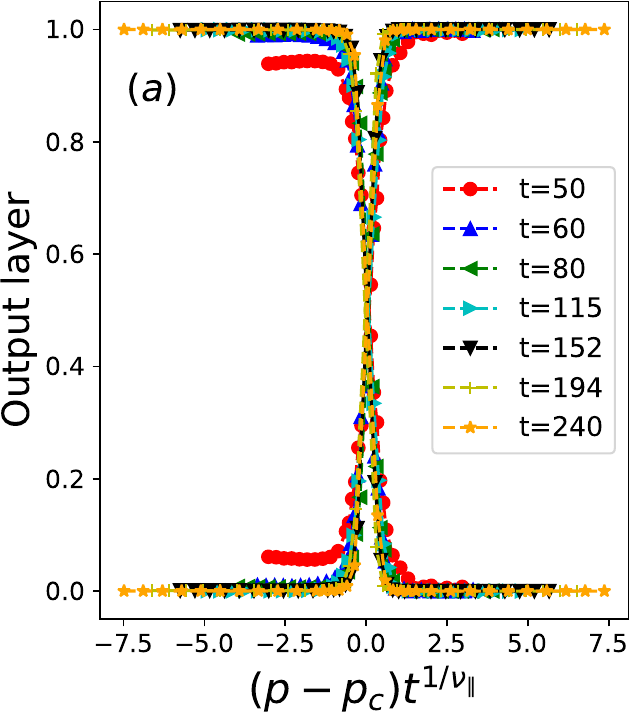} &
    \includegraphics[width=0.49\columnwidth]{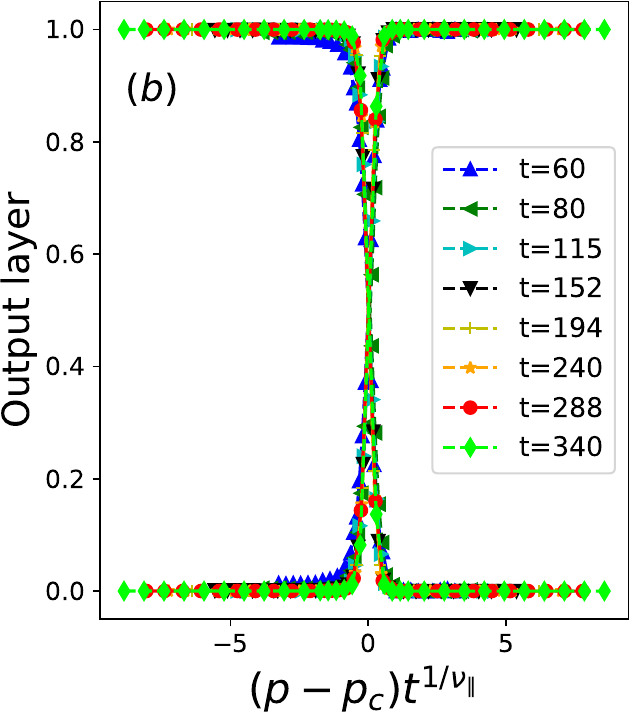} \\
    \includegraphics[width=0.49\columnwidth]{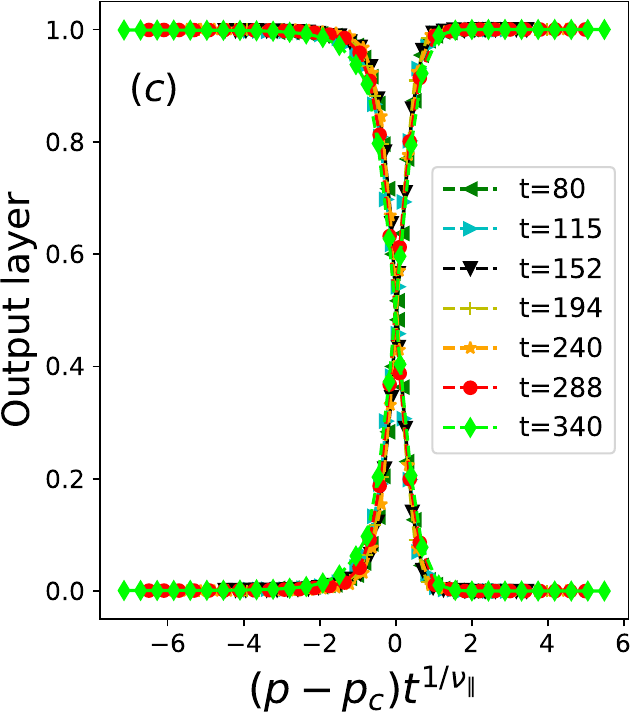} &
    \includegraphics[width=0.49\columnwidth]{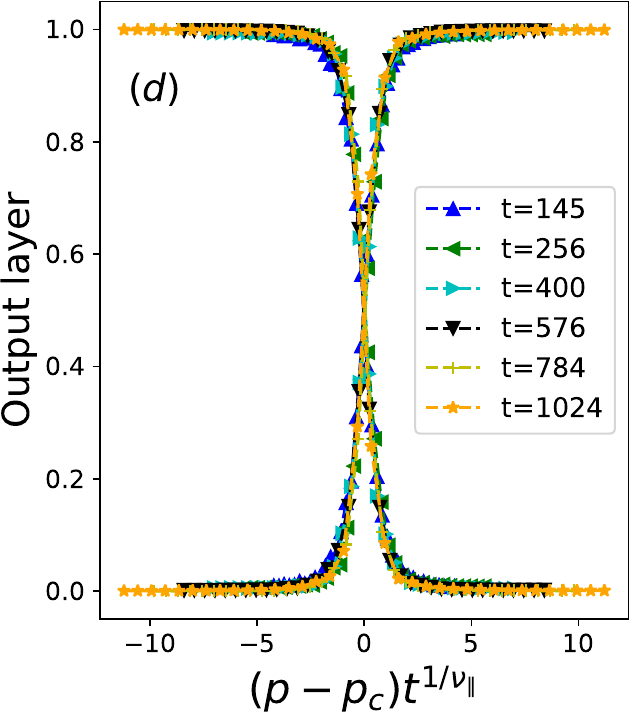}
    \end{tabular}
    \caption{CNN outputs results as a function of $(p-p_c)t^{1/{\nu_{\Vert}}}$ for (a) the bond DP ($L=8$, 12, 16, 20, 24, 28, 32; $t=50$, 60, 80, 115, 152, 194, 240),  (b) the site DP ($L=12$, 16, 20, 24, 28, 32, 36, 40; $t=60$, 80, 115, 152, 194, 240, 288, 340), (c) the Wolfram rule 18 ($L=16$, 20, 24, 28, 32, 36, 40; $t=80$, 115, 152, 194, 240, 288, 340), (d) the CDP ($L=12$, 16, 20, 24, 28, 32; $t=145$, 256, 400, 576, 784, 1024).}
    \label{fig:collapse2}
\end{figure}

Before we proceed, let us note that the bond DP, the site DP, and the Wolfram rule 18 of the (1+1)-dimensional DK all belong to the (1+1)-dimenional DP universality class,  which is characterized by the correlation exponent $\nu_\bot\approx 1.0968(4)$ and the temporal correlation exponent $\nu_\Vert \approx 1.7338(6)$ \cite{jensen1996low}. The CDP represents a different universality class (the CDP universality class) of absorbing phase transitions, where the percolation clusters are compact objects, which is characterized by $\nu_\bot=1$ and $\nu_\Vert= 2$ in $1+1$ dimensions\cite{domany1984equivalence,essam1989directed}.

Now, rescaling the probability $p-p_c$ by choosing proper $\nu_\bot$ and $\nu_\Vert$ in Fig.~\ref{fig:supervised} should render the output layer curves for different sizes collapse to the scaling functions $\hat{P}_{0|1}(x)$ and $\hat{P}_{0|1}(y)$. As seen in Fig.~\ref{fig:collapse1}, the curves coincide for different sizes with a suitable choice of ${\nu}_\bot$. The estimated DK critical exponents are ${\nu}_\bot=1.09\pm 0.02$ for the bond DP, ${\nu}_\bot=1.08\pm 0.03$ for the site DP, and ${\nu}_\bot=1.07\pm 0.03$ for the Wolfram rule 18, which are consistent with the theoretical value ${\nu}_\bot=1.0968$.  For the CDP, We estimate ${\nu}_\bot=0.99\pm 0.02$, which is again consistent with the theoretical value ${\nu}_\bot=1$. Similarly, Fig.~\ref{fig:collapse2} shows the data-collapse results for temporal correlations with respect to different simulation times. The estimated DK critical exponents are $\nu_\Vert=1.726\pm 0.02$ for the bond DP, $\nu_\Vert=1.723\pm 0.03$ for the site DP, and $\nu_\Vert=1.718\pm 0.03$ for the Wolfram rule 18, which are consistent with the theoretical value $\nu_\Vert=1.7338(6)$. We also estimate $\nu_\Vert=1.965\pm 0.03$ for the CDP, agreeing  with the theoretical value $\nu_\Vert=2$. According to $z=\nu_\Vert/ \nu_\bot$, the estimated DK dynamical exponents are $z=1.583\pm 0.02$ for the bond DP, $z=1.595\pm 0.03$ for the site DP, $z=1.606\pm 0.03$ for the Wolfram rule 18, $z=1.985\pm 0.03$ for the CDP.

Machine learning has been well applied to studying equilibrium phase transition models, but applying it to studying nonequilibrium phase transitions is a new research field, having attracted much attention in recent years. Previously, it has been demonstrated that spatial correlation exponents of nonequilibrium phase transitionos can be extracted in a similar manner as in the nonequilibrium case \cite{Li}.  Here, we explore further and find that the CNN output layer also contains temporal correlation information, which permits the extraction of temporal correlation exponents.

\section{semi-supervised learning of the Domany-Kinzel model \label{sec:DK semi-supervised}}

Semi-supervised learning is a kind of learning paradigm which combines supervised learning with unsupervised learning. The goal of semi-supervised learning is to obtain a predictive model by utilizing both labelled and unlabelled data in the training set, in which the unlabelled data in the training set become pseudo-labelled through the partially trained model, which is further updated by combining the original labelled data and these psedo-labelled data \cite{zhu2022introduction,berthelot2019mixmatch}. In this way, one may substantially reduce the cost for data labelling, or, for more practical applications, one only needs to label data with the most certainty and leaves those less certain ones unlabelled in the training set.

For the DK model, simulations are run on arrays of size $L=16$, up to $t=120$ steps. For each probability, 2000 labelled configurations are generated for the training set and another 1000 configurations for the test set.
In semi-supervised learning, basically the same convolutional neural network as illustrated in Fig.~\ref{fig:CNN} is being used, only that the output layer contains only \textit{one neuron}. Instead of labelling the raw configuration data according to the phase they are in, raw configurations are labelled by their particle number densities $\rho$, inferred from the MC configurations directly. However, to test the ability of semi-supervised learning, only half of the training set corresponding to a sparser probability selection [e.g.~$p=(0.1, 0.3, 0.5, 0.7, 0.9)$] are labelled with the particle number density $\rho_i(p)$ for the $i$-th sample, while the testing set includes data for all $p$ values [e.g.~$p=(0, 0.1, 0.2, 0.3,...,0.9, 1)$]. Once the CNN is fully trained with respect to both the labelled and pseudo-labelled data in the training set, the CNN output for the test set should predict the particle number density.

\begin{figure}[!htbp]
\setlength{\tabcolsep}{0pt}
\centering
\begin{tabular}{cc}
     \includegraphics[width=0.49\columnwidth]{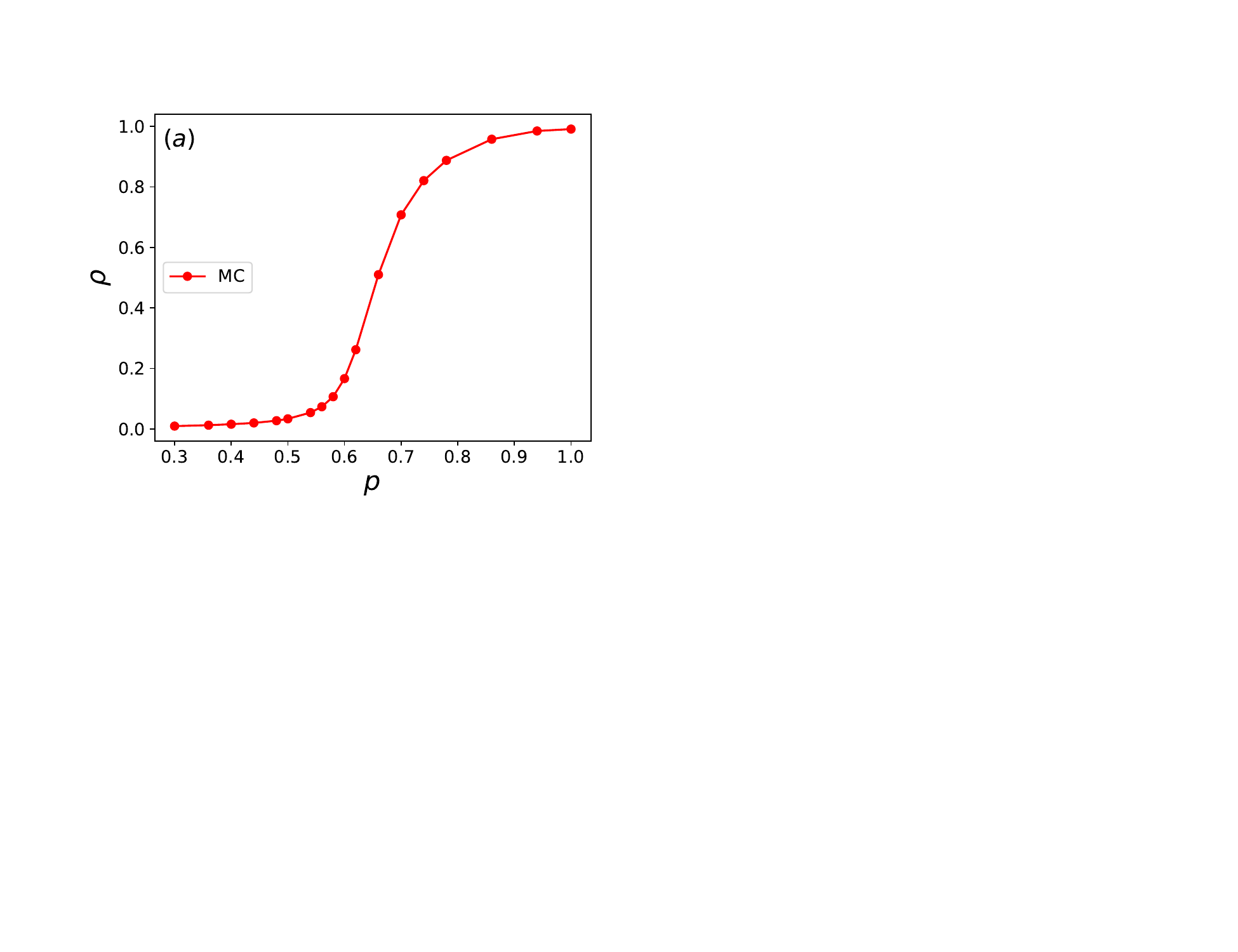} &
     \includegraphics[width=0.49\columnwidth]{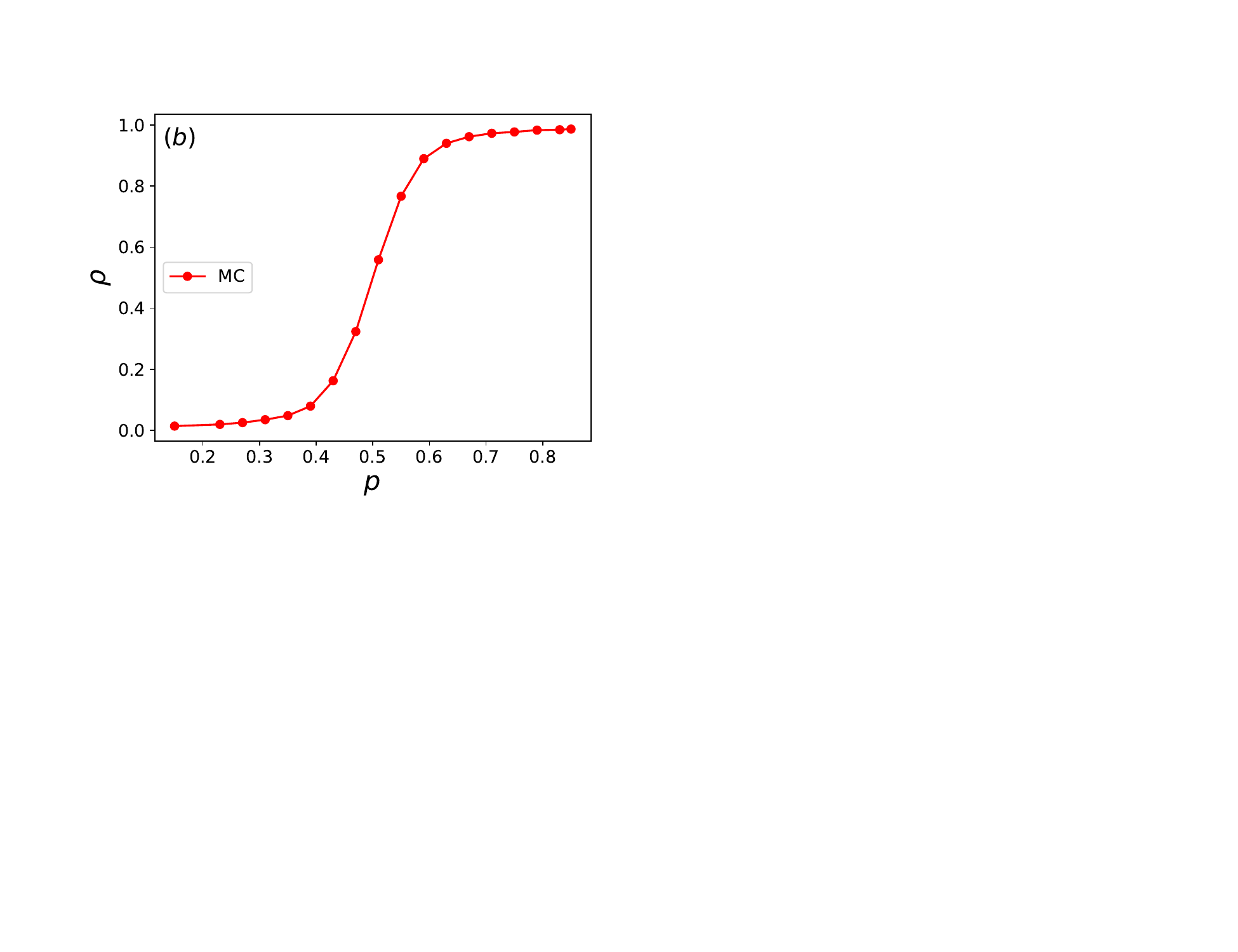} \\
     \includegraphics[width=0.49\columnwidth]{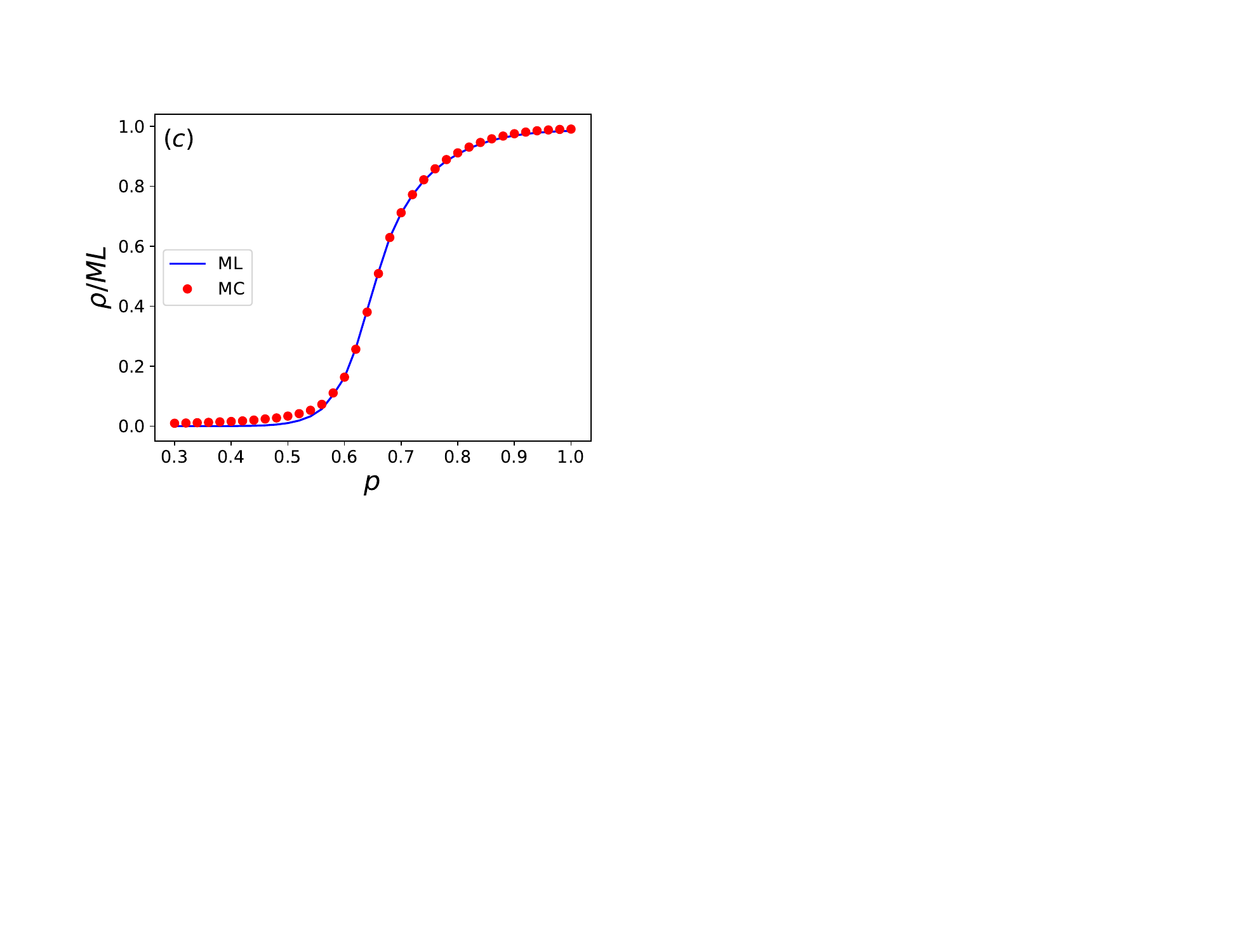} &
     \includegraphics[width=0.49\columnwidth]{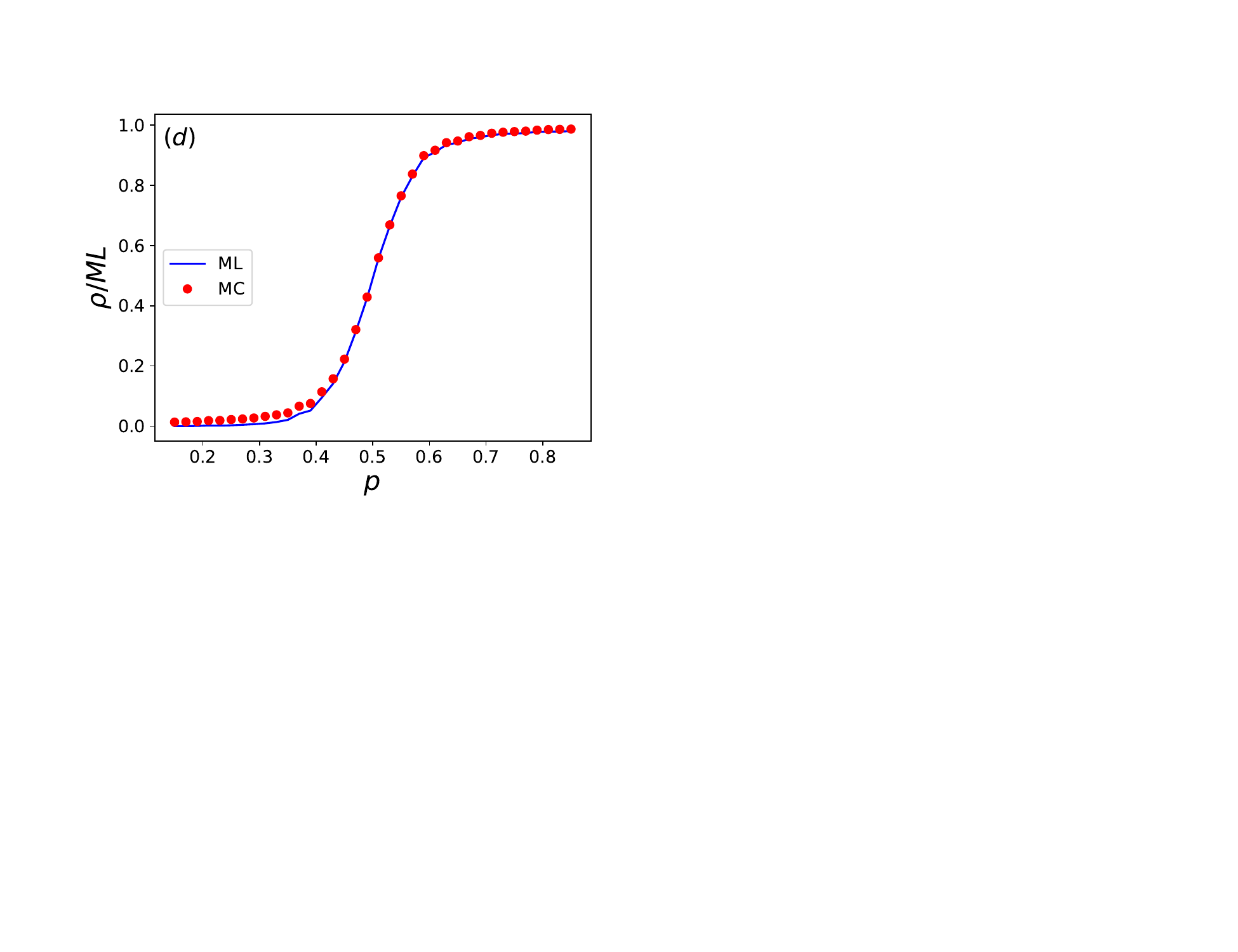} \\
     \includegraphics[width=0.49\columnwidth]{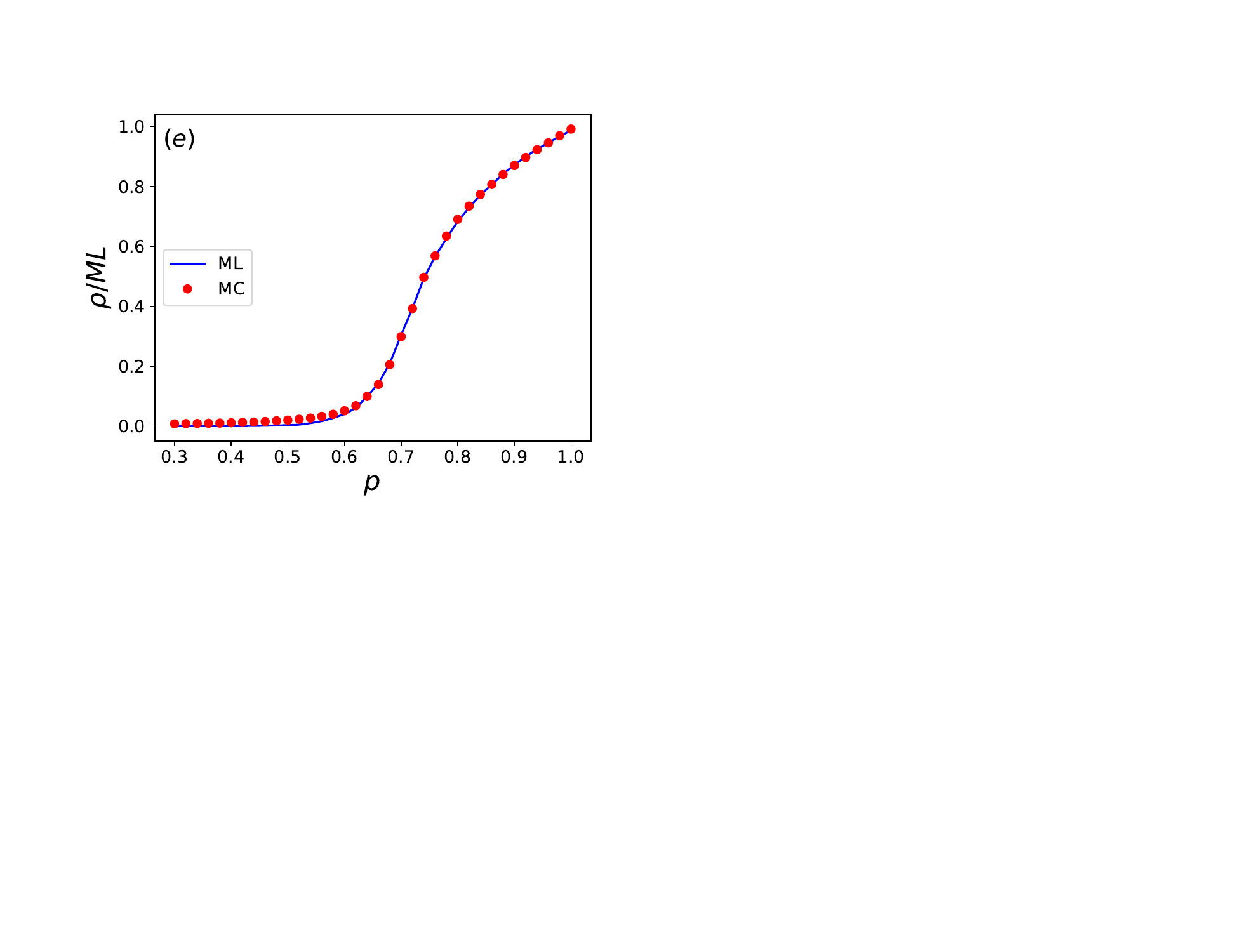} & \includegraphics[width=0.49\columnwidth]{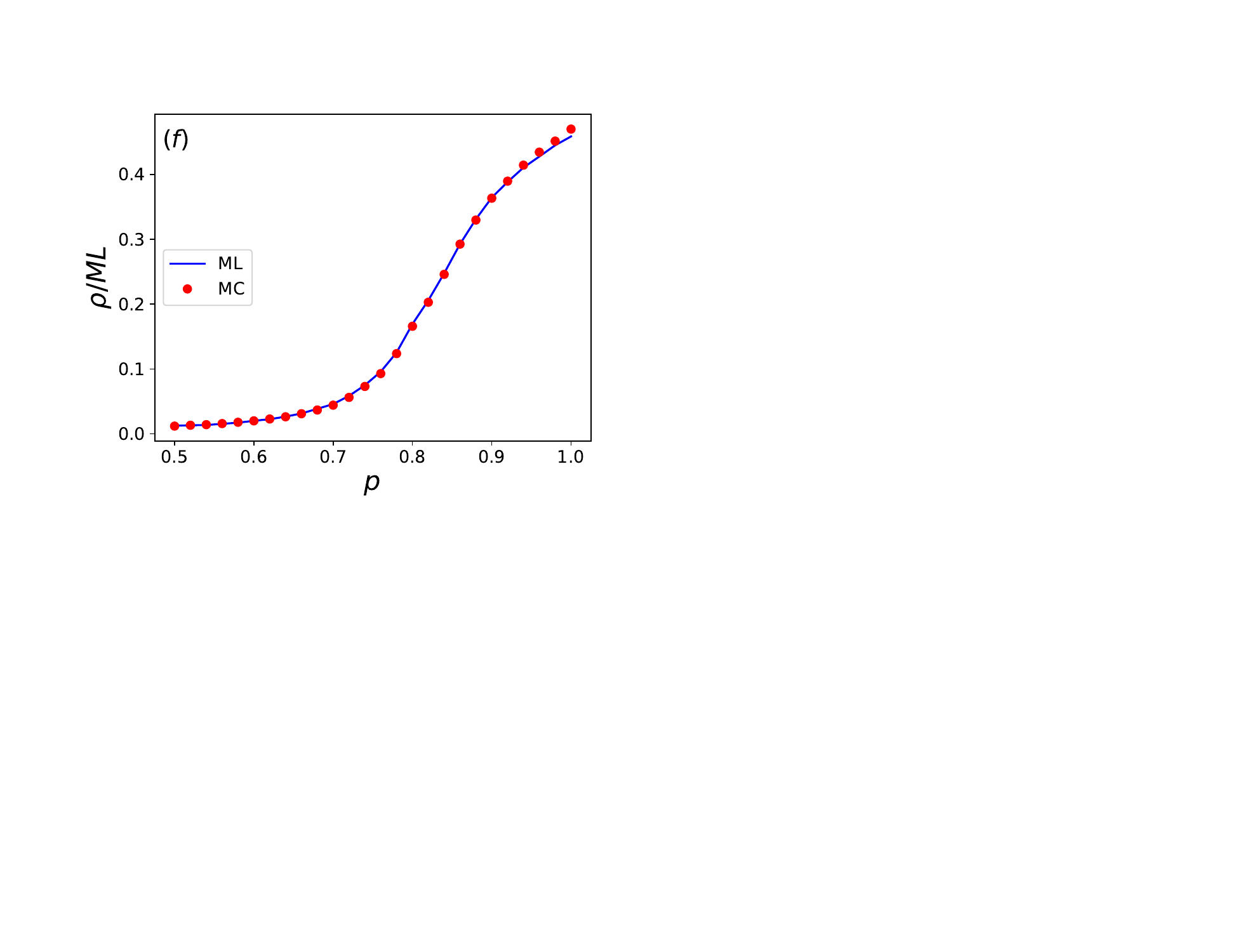}
\end{tabular}
\caption{Partially labelled training sets for (a) the bond DP and (b) the CDP. Semi-supervised learning of the particle number density with partially labelled training sets for (c) the bond DP, (d) the CDP, (e) the site DP, and (f) the Wolfram rule 18. Red dot represent particle number density of the DK model, blue line represent the CNN outputs. The predicted result is obtained after averaging over all the output results for the test set.}
\label{fig:semisupervised}
\end{figure}

The [Fig.~\ref{fig:semisupervised} (c)-(f)] show the results for the semi-supervised learning. We find that semi-supervised learning can predict the particle number density of the DK model, complying with the counterpart from the Monte Carlo simulations well. From the peak of the variance, the estimated values of critical points are $p_c=0.636\pm 0.02$ for bond DP, $p_c=0.695\pm 0.02$ for site DP, $p_c=0.495\pm 0.02$ for compact DP, and $p_c=0.792\pm 0.02$ for Wolfram rule 18.

 We remark that even for the DK model, the selection of the labelled portion in the training set could be quite arbitrary. This opens a possible avenue for the study of more intricate phase transitions such as topological phase transitions. Previously, it had been demonstrated that the unsupervised learning methods (PCA) are not suitable for extracting the critical points of the XY model \cite{hu2017discovering}. It would be interesting to study if one can infer the full phase information for topological phase transitions via semi-supervised learning by utilizing only partial information of these transitions.

\section{Unsupervised learning of the Domany-Kinzel model \label{sec:DK Unsupervised}}

In certain scenario, there may be a total absence of category information for the interested data (unlablled data) and unsupervised learning methods that realize sample classification through data analysis then become indispensable. In this section, two well-known unsupervised learning methods, i.e.~autoencoder and Principal Component Analysis (PCA), will be applied to detect the phases of the DK model. 

\subsection{Autoencoder results of the DK model}

\begin{figure}[t]
  \centering
    \includegraphics[width=0.8\columnwidth]{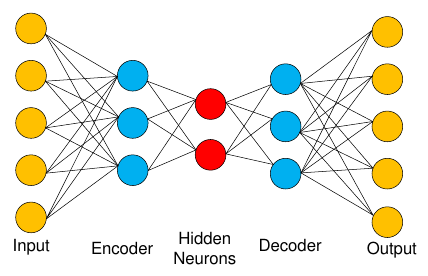}
    \caption{Schematic structure of Autoencoder. The fully connected autoencoder architecture  includes an input layer, an encoder, a latent layer of hidden neurons, a decoder, and an output layer. The autoencoder with number of neurons (1936, 968, 480, 240, 120, 60, 16, 2/1, 16, 60, 120, 240, 480, 968, 1936), relu activation functions.}
   \label{fig:Autoencoder}
\end{figure}

Autoencoders are simple generative models which can produce random outputs that are similar to the inputs \cite{liou2014autoencoder,ng2011sparse,bank2023autoencoders}. As illustrated in Fig.~\ref{fig:Autoencoder}, the fully connected autoencoder architecture that we used includes an input layer, an encoder, a latent layer of hidden neurons, a decoder, and an output layer. The inputs for the autoencoders are just raw DK configurations $x_i$. The model is trained until the L2 loss function
\begin{equation}
L(\phi,\theta)=\frac{1}{N}\sum_{i=1}^{N}||x_i-D_\theta(E_\phi(x_i))||^2_2
\label{eqs:encodeL}
\end{equation}
is minimized with respect to the encoder parameters $\phi$ and the decoder parameters $\theta$, with $E_\phi$ and $D_\theta$ being the encoder and the decoder functions, respectively.  In this way, an effective representation, that preserves the most important information of the input data, is achieved in the latent layer through data compression, which further permits data reconstruction via data decompression with the decoder.

For the DK model, simulations are run on arrays of size $L=16$, up to $t=120$ steps. For each probability $p$, 2000 configurations are generated for the training set, and another 1000 configurations for the test set. Note that the autoencoder output is limited to two hidden neurons,  meaning that the DK configurations are compressed into two dimensions. Once the autoencoder is trained, each input $\mathbf{x}_i$ from the test set then gives rise to one point $(h_{i1},h_{i2})$ on a two-dimensional plane corresponding to the degree of freedom of the latent layer.

\begin{figure}[!htbp]
\setlength{\tabcolsep}{0pt}
\centering
\begin{tabular}{cc}
     \includegraphics[width=0.45\columnwidth]{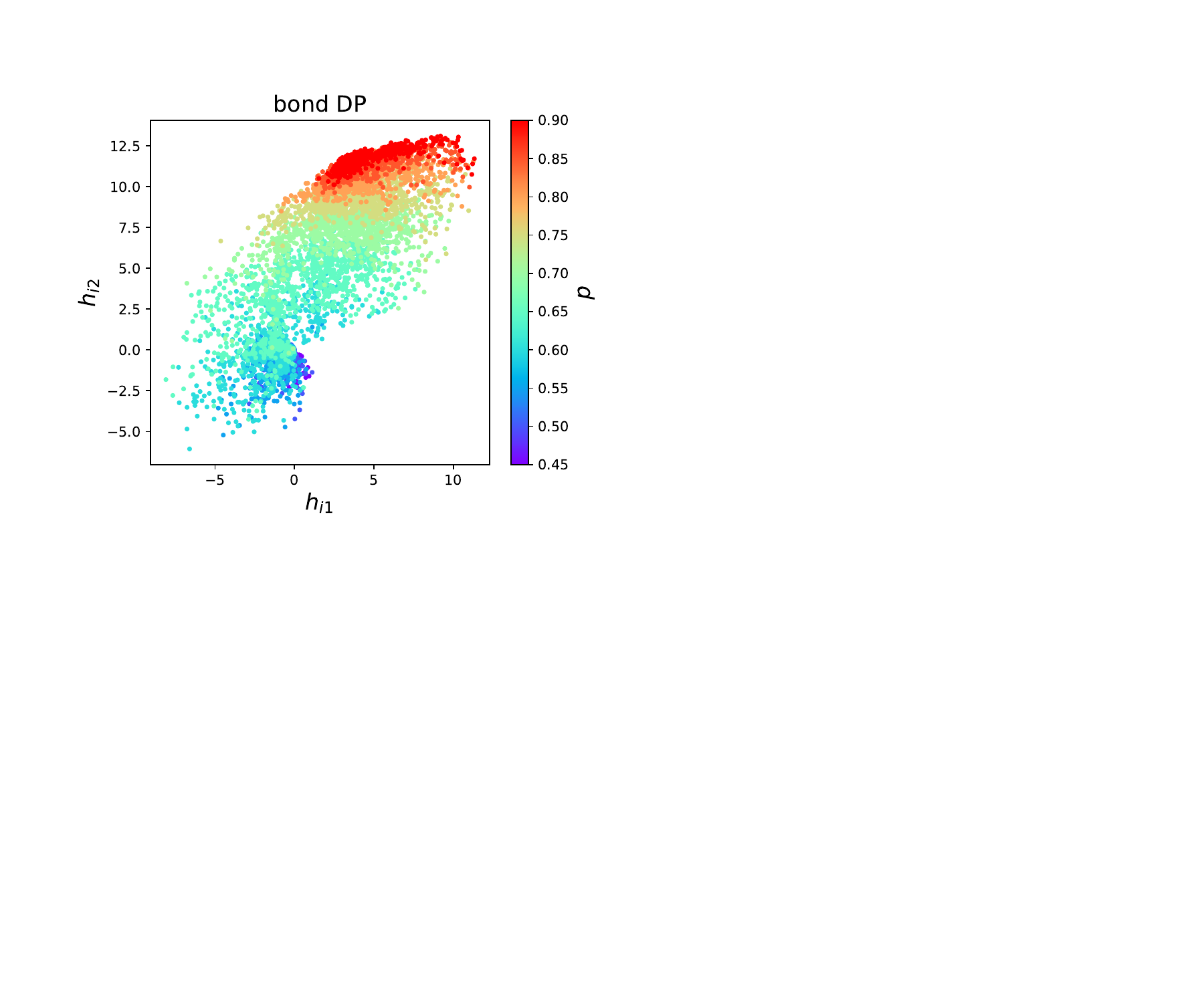} &
     \includegraphics[width=0.45\columnwidth]{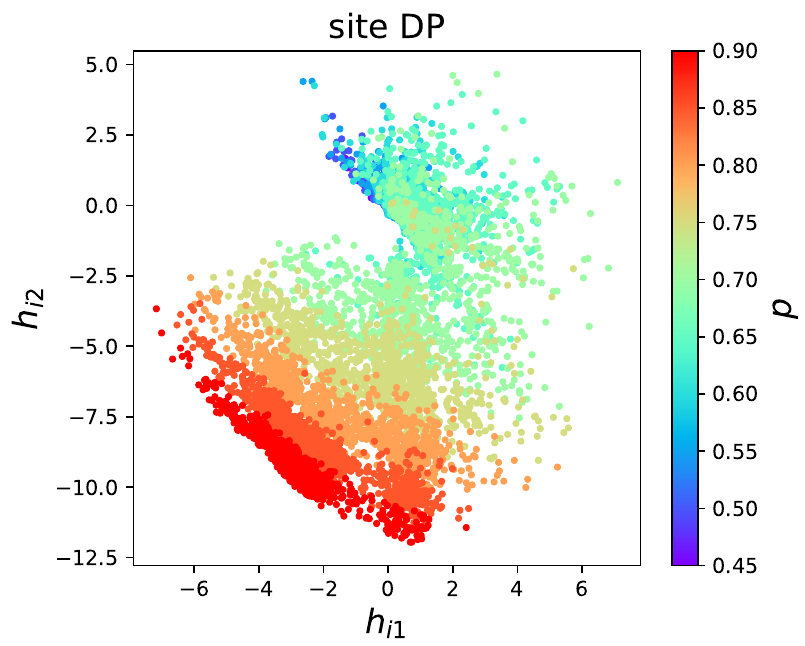} \\
     \includegraphics[width=0.45\columnwidth]{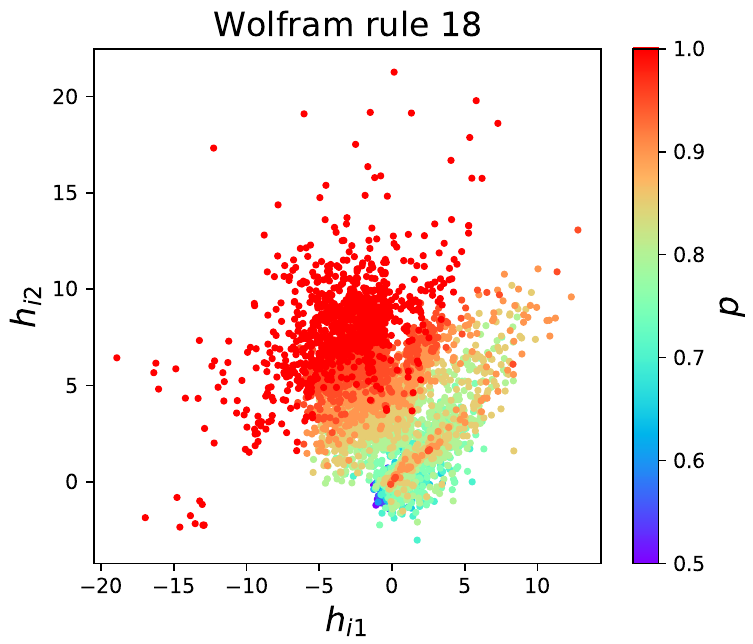} &
     \includegraphics[width=0.45\columnwidth]{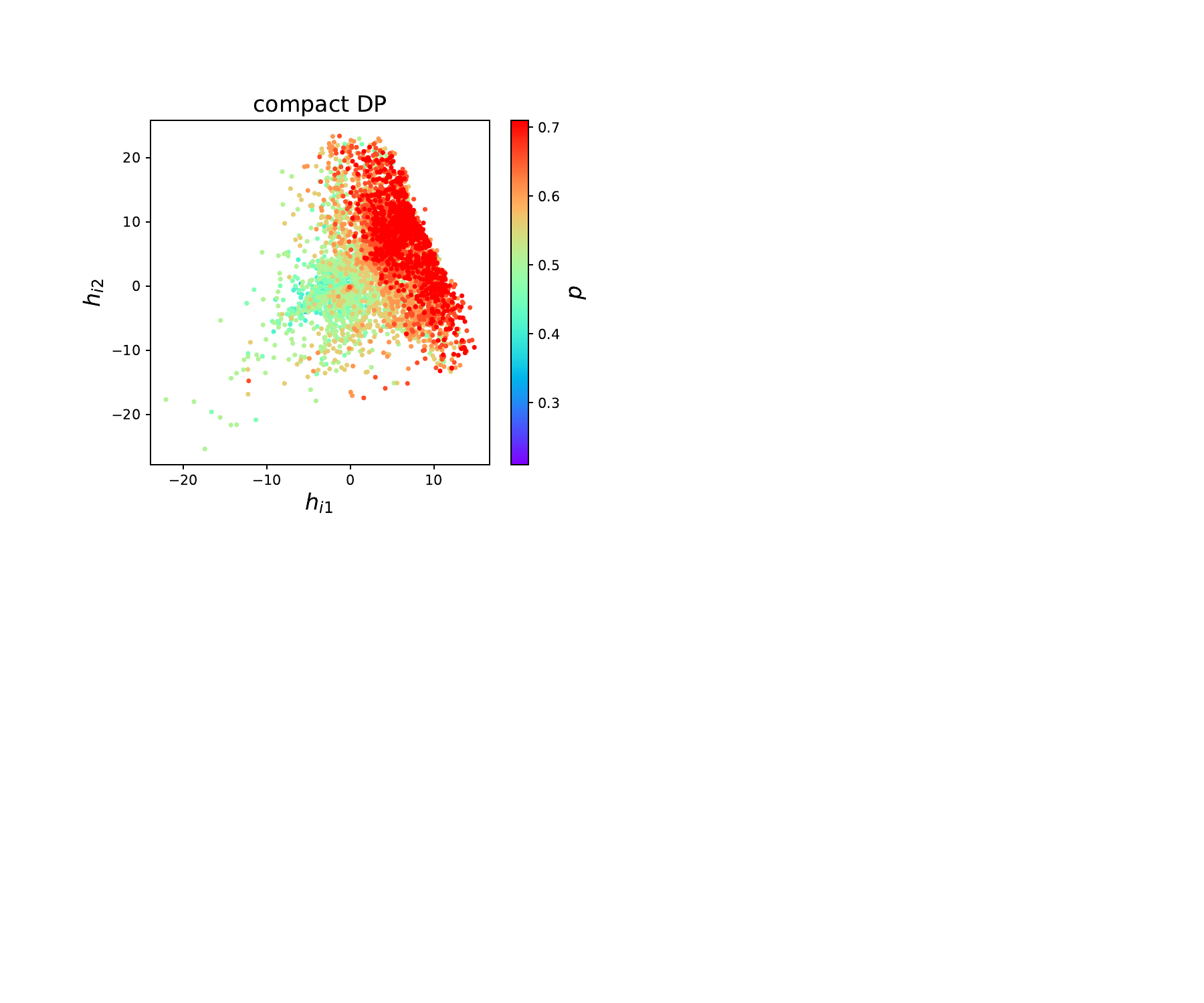}
\end{tabular}
\caption{The autoencoder output is linked with two hidden neurons, projecting the configurations of the bond DP, the site DP, the Wolfram rule 18, and the compact DP onto two dimensions. The colormap represent the probability $p$.}
\label{fig:Autoencoder_bond_site}
\end{figure}

As shown in Fig.~\ref{fig:Autoencoder_bond_site}, the used autoencoder roughly classifies the DK configurations into two clusters, although the absorbing phase and the active phase are not completely separated. Especially, configurations drawn from the same probability $p$ are closely clustered together, so the fuzzy boundary of the two phases means that the transition is of continuous type.    
 
Hence, while two neurons in the latent layer are capable of clustering the DK configurations into two phases. It suggests that autoencoders can capture essential information of the input data so as to detect the phases, without any prior knowledge of the DK model.

\subsection{PCA results of the DK model}
Principal Component Analysis (PCA) is also an unsupervised learning algorithm which can be used for data dimensionality reduction \cite{abdi2010principal,ringner2008principal,shlens2014tutorial} .  PCA performs orthogonal transformation on the data to find the direction of high variance, and converts the variables with possible correlations into linearly uncorrelated ones. The transformed variables are called principal components and here only the first two leading components will be used for the analysis of the DK model. One can intuitively imagine the process as projecting the data points of the original high-dimensional representation onto a lower dimensional space by selecting proper directions of projection with largest variances, so that the maximum amount of information is still preserved after reduction of the dimensions. Simulations are run on arrays of size $L=16$, up to $t=120$ steps. For each probability $p$, 1000 configurations are used to obtain the PCA results. 

\begin{figure}[!htbp]
\setlength{\tabcolsep}{0pt}
\centering
\begin{tabular}{cc}
     \includegraphics[width=0.49\columnwidth]{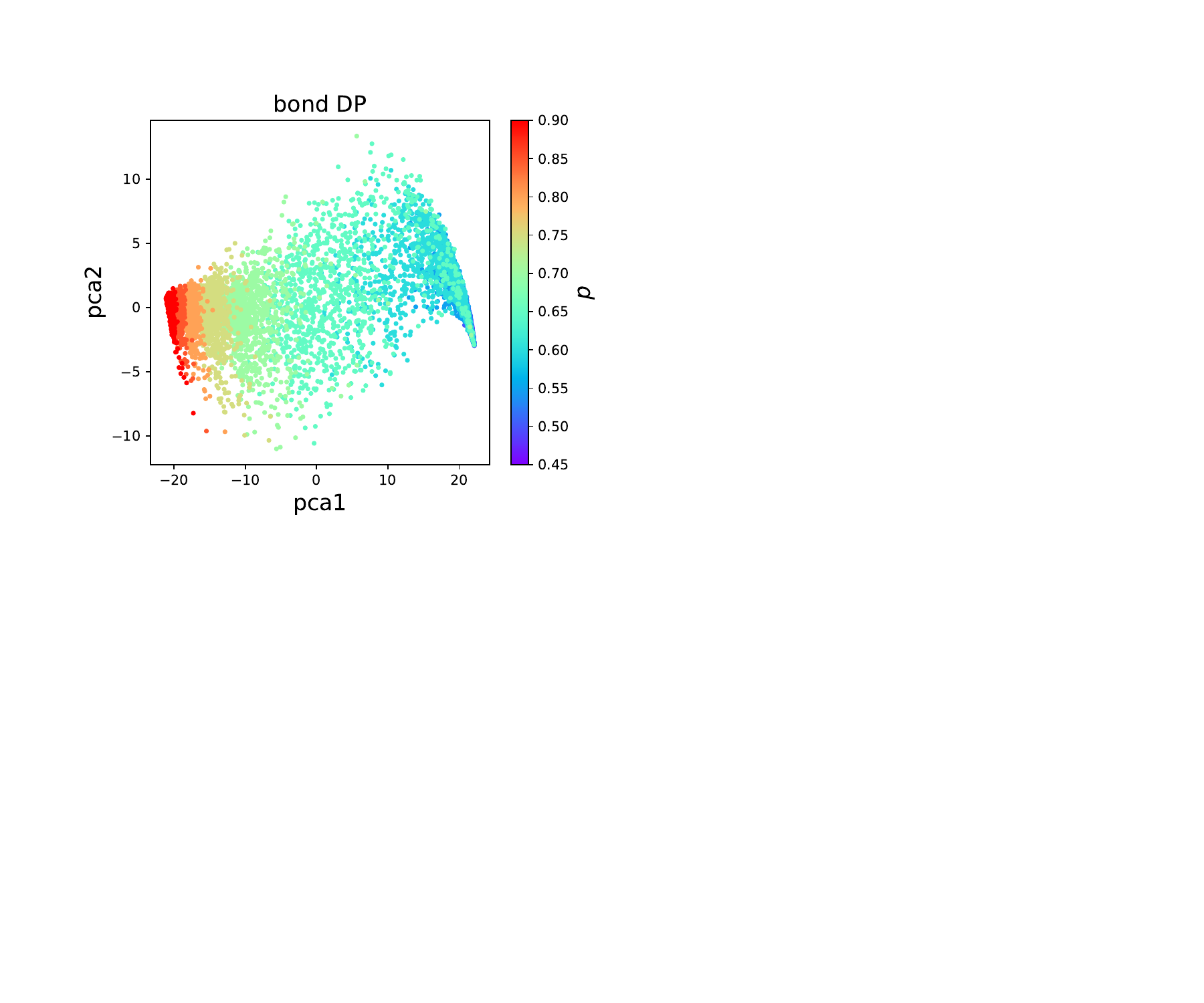} &
     \includegraphics[width=0.49\columnwidth]{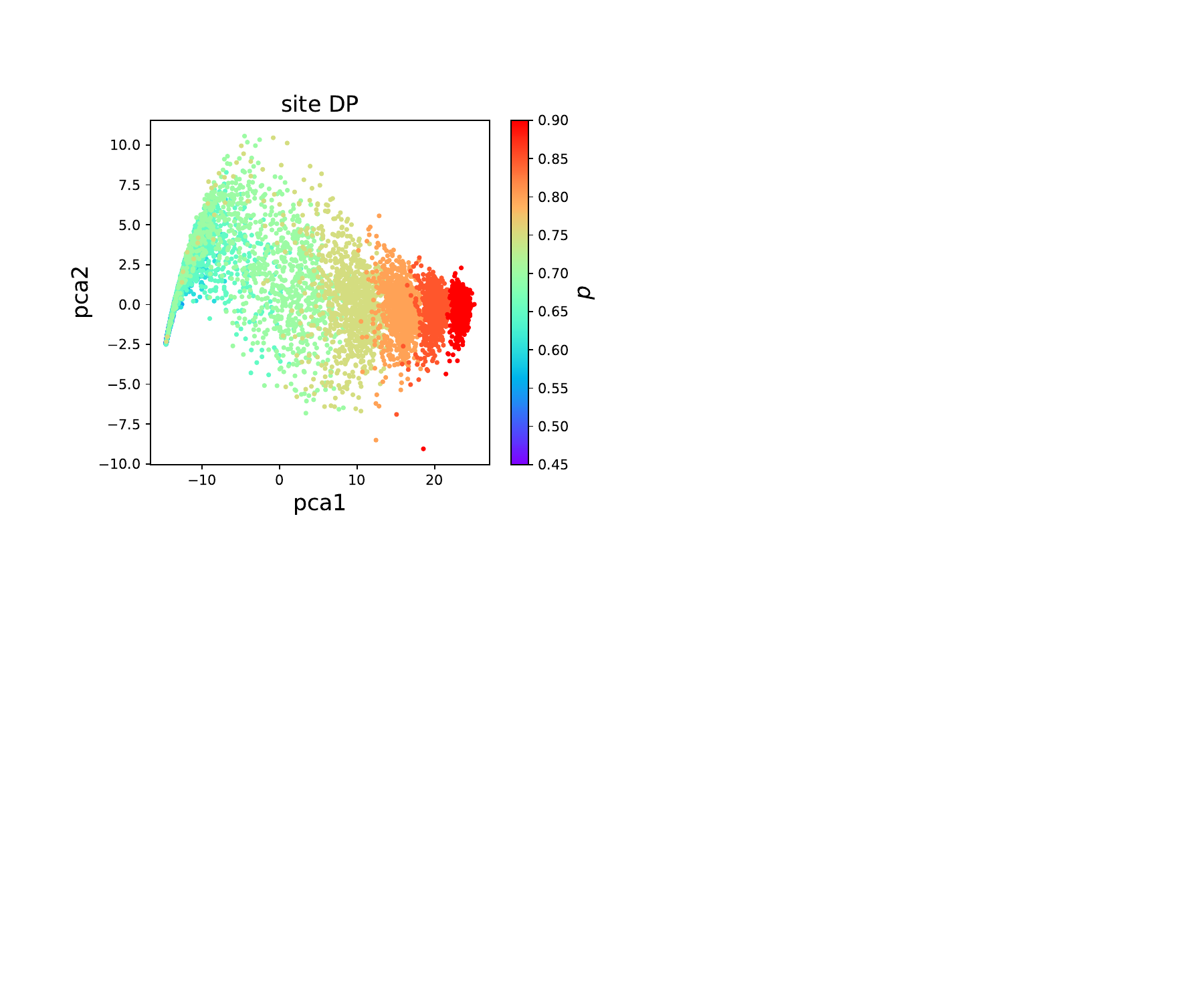} \\
     \includegraphics[width=0.49\columnwidth]{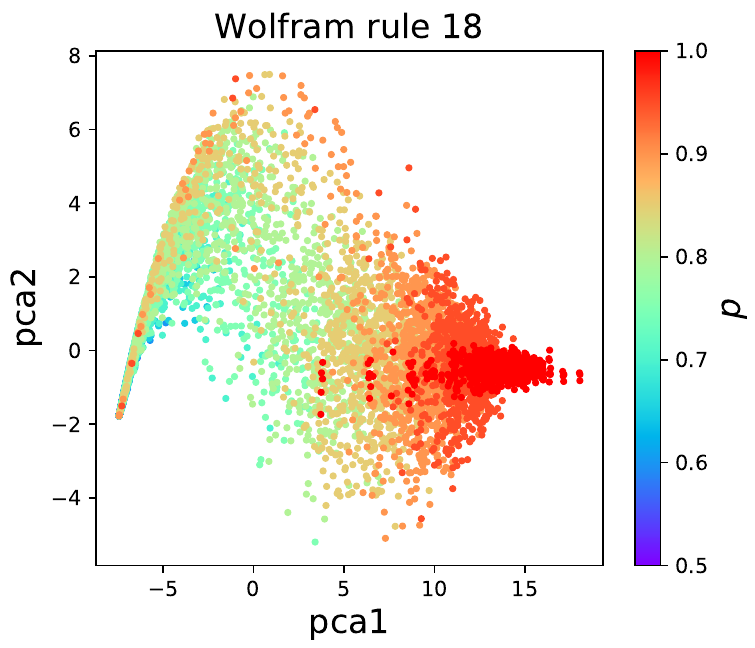} & 
     \includegraphics[width=0.49\columnwidth]{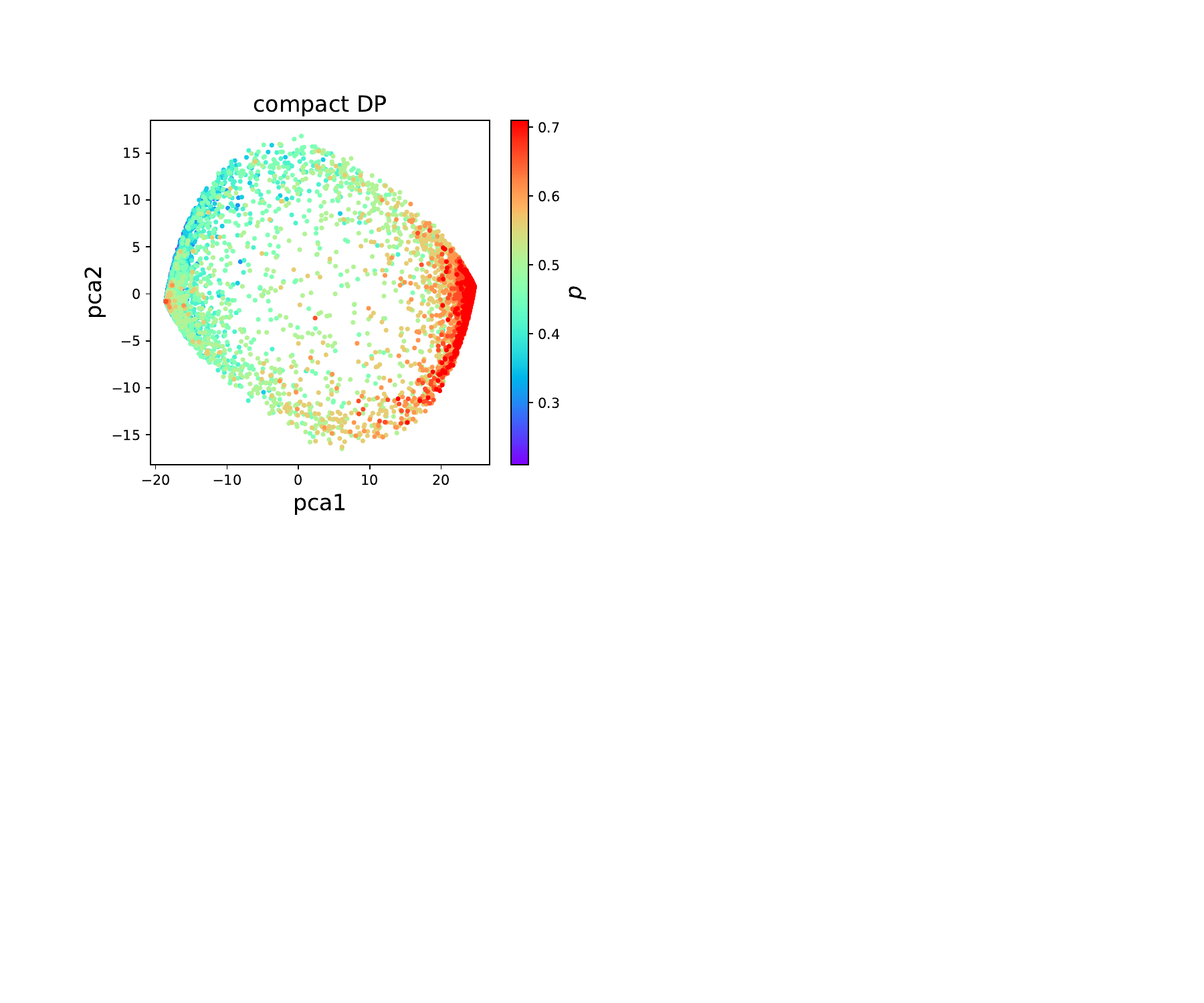}
\end{tabular}
\caption{PCA results for the bond DP, the site DP, the Wolfram rule 18, and the compact DP, 
with projection of the raw configurations onto the plane of the two leading principal components. The colormap represent the probability $p$.}
\label{fig:PCA_Bond_r}
\end{figure}

The two leading components of DK configurations by PCA are illustrated in Fig.~\ref{fig:PCA_Bond_r}. Similar to the autoencoder results, this reveals that PCA can also roughly classify the DK configurations into two phases. Since the boundary of these phases is not distinct, the transition is of continuous nature, as expected. 

Although PCA is just based on linear transformations of the
input data, we show that it is still effective in extracting features of the DK phase
transitions. Since the PCA could achieve similar results as compared to autoencoder without needing to train a model firstly, it costs less and is more convenient to apply.

\section{\protect Summary}
\label{sec:Summary}
In this paper, we applied supervised, semi-supervised and unsupervised learning methods to study the phase transitions and critical behavior of the Domany-Kinzel model. With supervised learning, the critical points were estimated from the neural network outputs. By further collapsing the outputs for different sizes, the correlation exponents $\nu _\bot$ and $\nu _\Vert$ were estimated, which are consistent with reference values in the literature. Previously, it has been demonstrated that, similar to the equilibrium case, the spatial correlation exponent $\nu _\bot$ of nonequilibrium phase transitionos can be extracted. Here, we explore further and find that the CNN output layer also contains temporal correlation information, which permits the extraction of the temporal correlation exponent $\nu _\Vert$. The achieved high accuracy values even for rather small system sizes suggest that the applied learning machine could learn the features of the phases quite well, so that the computation overheads in the MC simulation end can be substantially reduced.

The unsupervised learning methods, PCA and autoencoder, are able to roughly separate the phases into two clusters if the output is two-dimensional. Since learning through PCA is simpler, PCA generally is more efficient as compared to autoencoder. 

In semi-supervised learning, even though only half of the training set were labelled, by setting the output to just one neuron, we found that the network predicts the particle density of the test set quite well, permitting us to estimate the critical points of the DK model as well. Given these features of semi-supervised and unsupervised learning methods, it is advisable to use these methods to study more intricate phase transitions if data labelling becomes costly. 

Finally, we remark that even though only universal property of the DK model should matter along the DP critical transition line, the non-universal ``lacunarity'' property of clusters affects the learning accuracy. It is observed that learning machines generally learn the phase features of the DK models with denser clusters (e.g.~the bond DP) better than models with sparser clusters (e.g.~the Wolfram rule 18).

\begin{acknowledgments} 
We thank Jianmin Shen and Longfeng Zhao for valuable discussions. This work was partially supported by the Fundamental Research Funds for the Central Universities, China (Grant No.~CCNU19QN029), the National Natural Science Foundation of China (Grant No.~11505071, 61702207 and 61873104), and the 111 Project 2.0, with Grant No.~BP0820038.
\end{acknowledgments}

\bibliographystyle{apsrev4-2}
\bibliography{ADK}%
\end{document}